\documentclass[useAMS,usenatbib]{mn2e}
\usepackage{amssymb, amsmath, epsfig}
\newcommand{\bfk}{{\bf k}}

\newcommand{\nhat}{{\hat{\boldsymbol{n}}}}
\newcommand{\Mpc}{{\rm Mpc}}

\newcommand{\Msun}{{\rm M}_{\odot}}
\newcommand{\MHz}{{\rm MHz}}

\def\apj{ApJ}
\def\apjl{ApJL}
\def\apjs{ApJS}
\def\aj{AJ}
\def\mnras{MNRAS}

\begin{document}

\title{The Morphology of HII Regions during Reionization}

\title{The Morphology of HII Regions during Reionization}
\author[M. McQuinn et al.]{Matthew McQuinn$^1$\thanks{mmcquinn@cfa.harvard.edu}, 
Adam Lidz$^1$, 
Oliver Zahn$^{1,2}$, 
Suvendra Dutta$^{1}$,
Lars Hernquist$^{1}$,\newauthor
Matias Zaldarriaga$^{1,3}$\\
$^{1}$ Harvard-Smithsonian Center for Astrophysics, 60
Garden St., Cambridge, MA 02138\\
$^2$ Institute for Theoretical Astrophysics, University of
Heidelberg, Albert-Ueberle-Strasse 2, 69117 Heidelberg, Germany\\
$^3$ Jefferson Laboratory of Physics, Harvard University,
Cambridge, MA 02138\\
}

\pubyear{2006} \volume{000} \pagerange{1}
\maketitle\label{firstpage}

\begin{abstract}
It is possible that the properties of HII regions during reionization
depend sensitively on many poorly constrained quantities (the nature
of the ionizing sources, the clumpiness of the gas in the IGM, the
degree to which photo-ionizing feedback suppresses the abundance of
low mass galaxies, etc.), making it extremely difficult to interpret
upcoming observations of this epoch.  We demonstrate that the actual
situation is more encouraging, using a suite of radiative transfer
simulations, post-processed on outputs from a $1024^3$, $94$ Mpc
N-body simulation.  Analytic prescriptions are used to incorporate
small-scale structures that affect reionization, yet remain unresolved
in the N-body simulation.  We show that the morphology of the HII
regions for reionization by POPII-like stars is most dependent on the
global ionization fraction $\bar{x}_i$.  Changing other parameters by
an order of magnitude for fixed $\bar{x}_i$ often results in similar
bubble sizes and shapes.  The next most important dependence is on the
properties of the ionizing sources.  The rarer the sources, the larger
and more spherical the HII regions become.  The typical bubble size
can vary by as much as a factor of $4$ at fixed $\bar{x}_i$ between
different possible source prescriptions.  The final relevant factor is
the abundance of minihalos or of Lyman-limit systems.  These systems
suppress the largest bubbles from growing, and the magnitude of this
suppression depends on the thermal history of the gas as well as the
rate at which these systems are photo-evaporated.  We find that neither
source suppression owing to photo-heating nor small-scale gas clumping
significantly affect the large-scale structure of the HII regions,
with the ionization fraction power spectrum at fixed $\bar{x}_i$
differing by less than $20\%$ for $k < 5 ~\Mpc^{-1}$ between all the
source suppression and clumping models we consider.  Analytic models
of reionization are successful at predicting many of the features seen
in our simulations.  We discuss how observations of the 21cm line with
MWA and LOFAR can constrain properties of reionization, and we study
the effect patchy reionization has on the statistics of Ly$\alpha$
emitting galaxies.
\end{abstract}


\section{Introduction}
\label{intro} To interpret existing and future 
observations of the high redshift Universe, we need to understand the
morphology of the HII regions during reionization.  Observations of
high redshift quasars, gamma ray bursts, and Ly$\alpha$ emitters are
currently probing redshifts at which the Universe may have been
significantly neutral \citep{becker01, white03, fan06, totani05,
kashikawa06, santos04b}. Patchy reionization will leave its signature
in the spectra of quasars and gamma ray bursts \citep{haiman99,
miralda00, madau00, furlanetto04c} and in the correlation and
luminosity functions of Ly$\alpha$ emitters \citep{haiman02, santos04,
furl-galaxies05}.  Starting in 2007, the Atacama Cosmology Telescope
and the South Pole Telescope will dissect the high-$l$ CMB
anisotropies. The size distribution of HII regions during reionization
affects the spectrum of these anisotropies \citep{mcquinn05, zahn05a,
iliev06a}. Finally, 21cm maps of the reionizing Universe may soon
be available. The GMRT, LOFAR, and MWA arrays will begin observing
high redshift 21cm emission within the next few years.\footnote{For
more information, see http://www.lofar.org/, and
http://web.haystack.mit.edu/arrays/MWA/.}  The 21cm signal will be an
excellent probe of the structure of reionization \citep{zald04,
furlanetto04a, furlanetto04b, mellema06, furlanetto06b}.

A proper interpretation of these observations requires an
understanding of how properties of the ionizing sources, how gas
clumping, and how source suppression from thermal feedback impact the
size distribution of HII regions.  It is computationally demanding to
simulate reionization in large enough volumes to capture the
large-scale bubble morphology, and many previous numerical studies
simulated only a limited number of reionization scenarios, making it
difficult to isolate the impact of each of the numerous uncertain parameters.

We do not know which objects reionized the Universe, although it is
most likely that stellar sources produced the bulk of the ionizing
photons (e.g., \citet{wyithe03}). In this case, it is unclear whether
the ionizing photons were produced by the more numerous galaxies with
halo masses $m \approx 10^8 \; M_{\odot}$ or mainly by rarer, more massive
galaxies. Locally, the rate at which dwarf galaxies convert gas into
stars scales as galaxy mass to the two-thirds power
\citep{kauffmann03}. If the same is true in the high redshift
Universe, then the more massive galaxies could dominate the production
of ionizing photons.  However, it might be easier for ionizing photons
to escape into the inter-galactic medium (IGM) from smaller galaxies
\citep{wood99}. Analytic models predict larger HII regions in
scenarios in which the most massive galaxies produce more of the
ionizing photons \citep{furl-models}. In spite of our ignorance
regarding which sources reionized the Universe, numerical studies have
yet to examine how reionization depends on the properties of ionizing
sources.

Further, we have little observational handle on the amount of
small-scale structure, or `gas clumping', in the high redshift IGM,
and researchers have not reached a consensus regarding its impact on
the morphology of reionization.  Many previous large-scale
reionization simulations have either entirely ignored structure on
scales smaller than the simulation grid cell or, despite inadequate
resolution, have incorporated it via a subgrid clumping factor
calculated from their large volume simulations \citep{sokasian03,
ciardi03-sim, iliev05, zahn06}.  Recently, there has been some effort
to calibrate subgrid clumping factors from an ensemble of small-box
simulations \citep{mellema06, kohler05}.  However, even these efforts
are very simplified.  No study has tried to isolate the effect that
gas clumping has on the size distribution and morphology of HII
regions.  If the morphology is very sensitive to
this clumping, it would be hard to trust the results from simulations.

Another relevant piece of physics is thermal feedback from
photo-heating the IGM, which can suppress star formation and
potentially alter the morphology of reionization.  However, the extent
to which the structure of reionization is affected by such feedback
has yet to be adequately addressed. \citet{kramer06}, utilizing an
analytic model for reionization that includes feedback (albeit, on
halos that cool via molecular line emission), found that it can have a
dramatic impact on bubble sizes, in some cases creating a bimodal
bubble size distribution.  Similar claims may also hold for thermal
feedback on galaxies that cool via atomic transitions -- the more
likely culprit to ionize the Universe. \citet{iliev06b} found using
radiative transfer simulations that thermal feedback plays a key role
during reionization, marginalizing the contribution from halos with
masses below $10^9 ~M_{\odot}$.

In addition, the presence of minihalos and the rate at which the gas
from these halos is photo-evaporated may shape reionization.
\citet{iliev-mini} show that a significant fraction of the ionizing
photons will be consumed by minihalos and claim that the effect of
minihalos on the morphology of reionization is similar to changing the
efficiency of the sources.  On the other hand, \citet{furlanetto05}
argue analytically that minihalos can create a well defined peak in
the bubble size distribution that is set by the mean free path for an
ionizing photon to be absorbed by a minihalo. The effect of minihalos
on the characteristics of the HII bubbles has not been
investigated in simulations.

In this paper, we present a suite of parameterized models, using large
volume radiative transfer simulations, to understand the impact of each
of these uncertain quantities on the morphology of reionization.
Realistic simulations of reionization require extremely large volumes
with high mass resolution.  Previous estimates suggest that, in order
to capture a representative sample of the Universe during
reionization, one needs a simulation box with a side length of
approximately $100 \, {\rm Mpc}$ comoving \citep{barkana03,
furlanetto04a}.  To resolve halos at the atomic hydrogen cooling mass
($m_{\rm cool} \sim 10^8 \, M_{\odot}$ at $z = 8$) in a simulation of this
volume, one needs about $30$ billion particles -- larger than any N-body
simulation to date.  In order to get around this computational
difficulty, we employ a hybrid scheme that combines a $1024^3$
particle, $94 \; \Mpc$ N-body simulation with a Press-Schechter merger
history tree.  The merger tree allows us to incorporate halos that are
unresolved in our N-body simulation.  Additional effects such as thermal
feedback and minihalo evaporation are incorporated in our simulations
with analytic prescriptions.

This paper is organized as follows.  In \S \ref{simulations} we
outline the N-body and radiative transfer codes used in this study.
The radiative transfer code is discussed in more detail in \S
\ref{code}.  Section \S \ref{mergertree} describes our method for
including unresolved low-mass halos.  In \S \ref{sources} we
investigate the effect of different source prescriptions on
reionization, and in \S \ref{feedback} we discuss the effect of source
suppression owing to photo-heating. Section \S \ref{recombinations}
considers the role of quasi-linear gas clumping and minihalos in
shaping the morphology of reionization.  Section \S \ref{redshift}
discusses the dependence of the morphology on the redshift of
reionization.  The relevance of the previous results to observations
of Ly$\alpha$ emitters and of high redshift 21cm emission are discussed in \S
\ref{observations}.

Throughout this paper we use a $\Lambda$CDM cosmology with $n_s = 1$,
$\sigma_8 = 0.9$, $\Omega_m = 0.3$, $\Omega_{\Lambda} = 0.7$,
$\Omega_b = .04$, and $h = 0.7$ \citep{spergel03}.  All distances in
this paper are in comoving units.

More recent measurements suggest that $\sigma_8$ may, in fact, be
lower than the value assumed in this work \citep{spergel06}.  The best
fit WMAP value is $\sigma_8 = 0.74 \pm .05$ and when combined with
other CMB experiments, the 2Df galaxy survey and the Ly$\alpha$ forest
becomes $\sigma_8 = 0.78 \pm .03$ \citep{viel06}.  A lower $\sigma_8$
reduces the number of ionizing sources during reionization.  However,
according to analytic models for the halo distribution, the sources in
a $\sigma_8=0.8$ universe are equivalent to those in a $\sigma_8=0.9$
universe at a slightly earlier time. Specifically, structure formation
in a $\sigma_8=0.8$ universe at redshift $1+z$ should be identical to
that in a $\sigma_8=0.9$ universe at $1+z^\prime = 9 \, (1+z)/8$.  This
occurs because halo abundances depend on $\sigma_8$ through the
combination $D(z) \, \sigma_8$, where $D(z) \sim 1/(1 +z)$ is the high
redshift growth factor.  Analytic models for reionization based on the
excursion set formalism also depend on $\sigma_8$ only through the
same combination $D(z) \, \sigma_8$.  Therefore, if $\sigma_8$ is
lower, this is equivalent to a simple re-mapping of redshifts.
Furthermore, in \S \ref{redshift} we demonstrate that the bubble
structure (at fixed ionized fraction) is relatively independent of
redshift and hence $\sigma_8$.

This paper focuses on predicting the large-scale morphology of
reionization, rather than precisely {\em when} reionization happens.
Furthermore, we do not focus on understanding the morphology at times
when the global ionized fraction is near zero or near unity -- in both
limits, detailed modeling of the complex radiative, thermal and
chemical feedback processes is essential and challenging.  Instead, we
focus on intermediate ionization fractions.  In addition, we do not
discuss the evolution of the ionizing background or the part in $10^4$
neutral fraction within the bubbles. We leave such discussion to
future work.

\section{Simulations}
\label{simulations}  We run a $1024^3$ N-body simulation in a box of
size $65.6 ~h^{-1}\, \Mpc$ with the TreePM code L-Gadget--2
\citep{springel05} to model the density field.  Outputs are stored on
$50$ million year intervals between the redshifts of $6$ and $20$. A
Friends-of-Friends algorithm with a linking length of $0.2$ times the
mean inter-particle spacing is used to identify virialized halos.

The simulated halo mass function matches the \citet{sheth02} mass
function for groups with at least 64 particles \citep{zahn06}.
However, the measured abundance of $32-64$ particle halos is below the
true value, but at an acceptable level. Thirty-two particle groups
correspond to a halo mass of $10^9 \, M_{\odot}$. Ideally, we would
like to resolve halos down to the atomic hydrogen cooling mass,
$m_{\rm cool} \approx 10^8 M_{\odot}$, which corresponds to the
minimum mass galaxy that can form stars.\footnote{The molecular
hydrogen gas cooling channel can lower the minimum galaxy mass. However,
Lyman--Werner photons from the first stars dissociate the molecular
hydrogen, probably eliminating this cooling channel prior to the time
when the Universe is significantly ionized \citep{haiman97}.}  We add
unresolved halos into the radiative transfer simulation using the
prescription described in \S \ref{mergertree}.

To generate the density grids, we use nearest grid point gridding of
the N-body particles.  Nearest grid point is problematic if Poisson
fluctuations in the number of particles are important at the cell
scale.  However, a typical cell in our fiducial runs has $64$ dark
matter particles, such that Poisson fluctuations are much smaller than
the order-unity cosmological ones at the cell scale.  Nearest grid
point affords us a higher level of gas clumping (and a more accurate
level of recombinations) than other gridding procedures, which smooth
the N-body density field more severely.

We use an improved version of the \citet{sokasian01} radiative
transfer code, which is discussed in detail in \S \ref{code}.  This
code is optimized to simulate reionization, making several justified
simplifications to drastically speed up the computation compared to
other reionization codes. The code inputs the particle locations from
the N-body simulation as well as a list of the ionizing sources, and
it casts rays from each source to compute the ionization field. We
assume that the sources have a soft UV spectrum that scales as
$\nu^{-4}$ (consistent with a POPII initial mass function (IMF)). The
parameters we choose for the source luminosities, subgrid clumping,
and feedback are varied throughout this paper and are discussed in
subsequent sections.

The radiative transfer code assumes perfectly sharp HII fronts,
tracking the front position at subgrid scales.\footnote{This is not
true for self-shielded regions, which can remain neutral behind the
front (see \S \ref{minihalos}).} This is an excellent approximation
for sources with a soft spectrum, in which the mean free path
for ionizing photons is kiloparsecs, substantially smaller
than the cell size in our radiative transfer simulations.

The radiative transfer simulations in this paper typically take two
days on a $2.2$ GHz AMD Opteron processor to reach an ionized fraction
of $\bar{x}_i = 0.8$.  We do not discuss ionization fractions larger
than $\bar{x}_i = 0.8$ in this work because our simulation box becomes
too small to provide a representative picture at larger $\bar{x}_i$.
In some models for reionization, our box is too small even at smaller
$\bar{x}_i$ than $0.8$ to adequately sample the bubble scale and
generate clean power spectra.

We typically choose source parameters so that reionization ends near
$z = 7$.  While overlap -- the final stage of reionization in which
the bubbles merge and fill all space -- may have occurred at higher
redshifts, upcoming observations of 21cm emission, QSOs, and
Ly$\alpha$ emitters are most sensitive to low redshifts reionization
scenarios. The most recent WMAP $\tau = 0.09 \pm 0.03$ is consistent
at the 1--$\sigma$ level with all the ionization histories in this
paper \citep{spergel06}.  Other papers have attempted to match the
source properties to observations at lower redshifts (e.g.,
\citet{gnedin00}).  The escaping UV luminosity of observed galaxies is
very uncertain, and current observations do not resolve low luminosity
galaxies at high redshifts. Significant extrapolation is hence
required to connect the properties of observed galaxies at lower
redshifts to the properties of the galaxies that reionize the Universe.
We expect that the source prescriptions adopted in this paper are
consistent with all current observational constraints.

Table \ref{table1} lists the parameters for the reionization
simulations discussed in this paper. A typical luminosity for a halo
of mass $m$ in the simulations is $\dot{N}(m)= 3 \times 10^{49} \;
m/(10^8 M_{\odot})$ ionizing photons $s^{-1}$. A Salpeter IMF yields
approximately $1.5 \times 10^{53}$ ionizing photons $s^{-1}$ yr
$M_\odot^{-1}$ \citep{hui02}. For an escape fraction of $ f_{\rm esc}
= 0.02$, for a Salpeter IMF, and for a typical $\dot{N}(m)$ in our
simulations, the star formation rate in a halo is
$\dot{S}(m) = m/(10^{10} \; M_{\odot}) ~ M_{\odot} ~ {\rm yr}^{-1}$.

\section{Unresolved Sources}
\label{mergertree} Our N-body simulation does not
resolve halos with masses less than $10^9 ~ \Msun$.  
We use an analytic prescription to include smaller mass halos that
are sufficiently massive for gas to cool by atomic processes and form stars.
It is unrealistic to ignore the effect of the halos with $m <
10^9 ~ \Msun$, as many previous studies have done, since these halos
contain more than half of the mass in cooled gas at all relevant
redshifts (modulo feedback from photo-heating). In addition, halos
smaller than the cooling mass can still affect the clumpiness of the
IGM and, thus, are important to incorporate in our simulations.

We outline two methods for adding unresolved halos to our simulation
in this section and discuss the merits of each method.  {\it Method
1}:  We add unresolved halos into each cell on the simulation mesh
according to the mean abundance predicted by Press-Schechter theory.
In this text, we use this method to include the minihalos. In a cell
of mass $M_c$ and linear overdensity today $\delta_{0,M}$, the
Press-Schechter mass function for halos with mass $m <M_c$ is
\begin{eqnarray}
n_{\rm PS}(m, \delta_{0, M}, M_c, z) &=& \sqrt{\frac{2}{\pi}} \,
\frac{\bar{\rho}}{m^2} ~\left | \frac{d \log \sigma}{d \log m}
\right | \frac{\delta_c(z) - \delta_{0, M}}{\sqrt{\sigma^2(m)-
\sigma^2(M_c)}} \nonumber \\
&\times& \exp \left[-\frac{(\delta_c(z) - \delta_{0, M})^2}{2[\sigma^2(m)
- \sigma^2(M_c)]}\right], \label{eq:psmf}
\end{eqnarray}
where the function $\sigma^2(M)$ is the linear-theory variance in a region of
Lagrangian mass $M$, $\bar{\rho}$ is the mean density of the Universe,
$\delta_c \approx 1.69/D(z)$, and $D(z)$ is the growth function
\citep{press, bond91}.  Halos cluster differently in Eulerian space,
and, to account for this, we relate the linear overdensity $\delta_0$
to the Eulerian space overdensity $\delta$ with the fitting
formula calibrated from spherical collapse \citep{mo96}:
\begin{eqnarray}
\delta_0 &=&\left[1.68647 - \frac{1.35}{(1 + \delta)^{2/3}} -
\frac{1.12431}{(1 + \delta)^{1/2}} + \frac{0.78785}{(1 +
\delta)^{0.58661}}\right] \nonumber\\
&\times& \frac{\delta_c(z)}{1.68647}.
\label{eqn:lag}
\end{eqnarray}

The radiative transfer code inputs the Eulerian overdensity $\delta_M$
for all cells from the N-body simulation.  To get the linear theory
overdensity $\delta_{0, M}$ we use equation (\ref{eqn:lag}) and
$\delta_M$.  In each cell of mass $M_c$ and linear overdensity
$\delta_{0, M}$, we place the average number of halos expected from
Press-Schechter theory using equation (\ref{eq:psmf}).  When including
the lower mass halos with this method, we need to choose a coarse cell
that contains more mass than the mass of our largest unresolved halo
or $10^9 ~ \Msun$.  We also need a scheme to distribute the halos
among the cells on the finer grid on which we perform the radiative
transfer. We discuss this scheme in \S \ref{minihalos}.

The disadvantage of Method 1 is that it involves putting the
average of the expected number of halos in each coarse cell and,
hence, ignores Poisson fluctuations in the halo abundance.  Even the
smallest galaxies at these high redshifts are rare and so Poisson
fluctuations in their abundance can be important.  {\it Method 2}: We
account for Poisson fluctuations by using the \citet{sheth99} merger
tree algorithm to generate the unresolved halos.  This algorithm
partitions a cell with mass $M_c$ into halos and, for a white noise
power spectrum, produces the correct average abundance of halos,
$n_{\rm PS}(m, \delta_{0,M}, M_c, z)$, as well as the correct
statistical fluctuations around this mean. The algorithm is guaranteed
to work only for a white noise power spectrum, but \citet{sheth99}
show that it works well at reproducing $n_{\rm PS}(m, \delta_{0,M}, M_c,
z)$ and other relevant statistics for more general power spectra.
This algorithm allows us to generate a spatially and temporally
consistent merger history tree. We find, for the small mass halos of
interest, that the algorithm generally produces more halos than the
Press-Schechter prediction. To compensate, we lower $\sigma_8$
slightly in the merger history computation to achieve the best
agreement with the Press-Schechter mass function for our fiducial
cosmology.

\begin{figure}
\rotatebox{-90}{ \epsfig{file=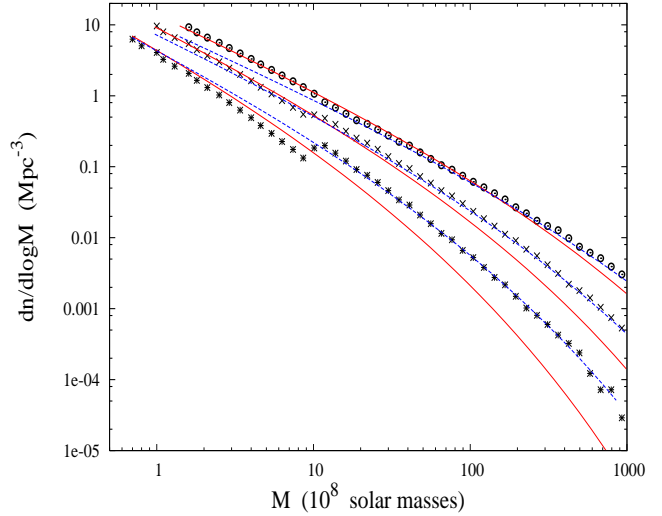, width=7cm,
height=8.6cm}}\caption{The halo mass function from our N-body
simulation and merger history tree algorithm. The merger history tree
generates all the halos below $10^9 ~ M_{\odot}$, and the more massive
halos derive from the N-body simulation.  We plot curves for the
merger tree plus resolved N-body halos at $z = 6.5$ ({\it circles}),
$8.7$ ({\it x's}) and $11.1$ ({\it asterisks}).  The solid curves are
the Press-Schechter, and the dashed are the Sheth-Tormen mass functions
for these redshifts.  The low-mass cutoff for these curves is set by
the HI atomic cooling mass $m_{\rm cool}$. The mass function of the
merger tree halos agree well with the theoretical curves (especially
at the lower two redshifts, which are the most relevant to this
study).} \label{fig:mf}
\end{figure}

Figure \ref{fig:mf} shows the halo mass function measured from our
simulations at $z = 6.5$ ({\it circles}), $8.7$ ({\it x's}) and $11.1$
({\it asterisks}).  The merger history tree generates halos below
$10^9 ~ M_{\odot}$, whereas the other, more massive halos are resolved
in the simulation.  The solid curves are Press-Schechter and the
dashed curves are Sheth-Tormen mass functions for these redshifts.
The mass function from the merger tree agrees best with
Press-Schechter and fairly well with Sheth-Tormen, particularly at
the lower two redshifts -- the most relevant redshifts for this study.
Note that the abundance of resolved halos in our simulation is closer
to the Sheth-Tormen mass function than to the Press-Schechter.

\begin{figure}
\rotatebox{-90}{ \epsfig{file=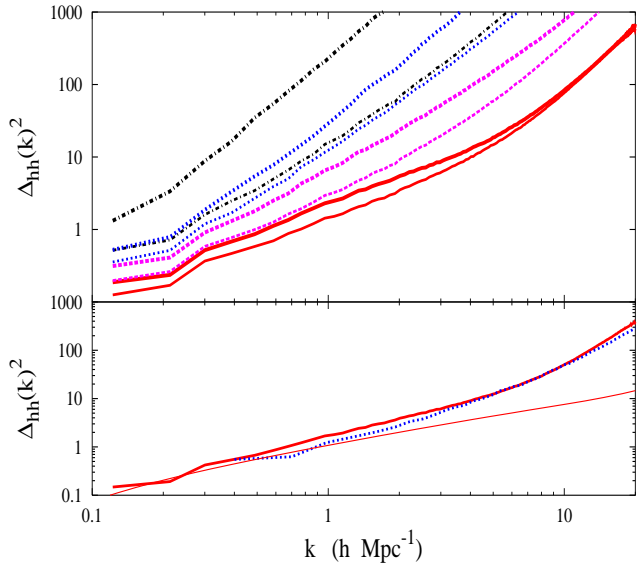, height=8.7cm, width=7.5cm}}
\caption{{\it Top Panel}: The mass-weighted halo power spectrum
$\Delta_{hh}^2$ for $z =6.6$ ({\it thin curves}) and $11.1$ ({\it
thick curves}) for several source models.  [$\Delta_{hh}^2 = k^3
\langle \delta \rho_h(k)^2\rangle/(2 \pi^2)$, where $\delta \rho_h(k)$
is the fluctuation in the halo mass density in Fourier space.] The
dot-dashed curves are for halos with $m >4\times10^{10} ~ M_{\odot}$,
the dashed are for the resolved halos with $m > 2 \times 10^9 ~
M_{\odot}$, and the solid are for all halos above the cooling mass.
We also include the power spectrum of halos above the cooling mass,
but weighted as $m^{5/3}$ rather than $m$ ({\it dotted curves}) to
match a source model discussed in \S \ref{sources}.  All halos with
masses between the cooling mass and $10^9 ~ M_{\odot}$ are
incorporated with the merger tree method.  {\it Bottom Panel}: The
mass-weighted power spectrum of the source halos (merger tree + resolved
halos) at $z = 8.7$ ({\it solid curve}), and the
power spectrum of a smaller box simulation ($20 \; h^{-1}~\Mpc$,
$1024^3$ particles) that resolves halos down to the cooling mass for
$z = 8.7$ ({\it dotted curve}).  The power spectrum from the merger
tree method agrees with this higher resolution run, boosting our
confidence in this method.  The thin solid curve is $k^3\,
\bar{b}_{PS}^2 \, P_{\delta \delta} (k, z)/2 \pi^2$, in which
$\bar{b}_{PS}$ is the average Press-Schechter bias and $P_{\delta
\delta}$ is calculated using the Peacock and Dodds fitting
formula. }
\label{fig:mtps}
\end{figure}

The merger history tree algorithm generates the halos in Lagrangian
space, requiring us to then map them to Eulerian space.  The
progenitor halos -- the halos at the lowest redshift bin such that
they sit on the top of a merger history tree -- are generated within
each coarse cell on a $64^3$ grid in Lagrangian space, and they are
then randomly associated with one of the fine cells within its
respective coarse cell (typically there are $256^3$ fine cells).  This
randomization is justified by the fact that Poisson fluctuations
dominate over cosmological fluctuations at the scale of the coarse
cell. To map our halos to Eulerian space in a self-consistent manner,
we associate each progenitor halo with a particle whose initial
(Lagrangian) position is the center of the same fine cell as the
Lagrangian position of the halo. We then displace the particle at each
redshift according to second order Lagrangian perturbation theory
\citep{crocce06}.  At higher redshifts, we split the progenitor halo into its
daughter halos, and all daughter halos are associated with the same
particle as their parent. This method for adding unresolved halos is
similar to the PT halo algorithm, an algorithm to quickly generate
mock galaxy surveys \citep{scoccimarro02}.

The bottom panel in Figure \ref{fig:mtps} plots the mass-weighted halo
power spectrum $\Delta_{hh}^2$ at $z = 8.7$ from a $1024^3$, $20
~h^{-1}$ Mpc box simulation that resolves halos down to the cooling
mass ({\it dotted curve}).  Note that $\Delta_{hh}^2 = k^3 \langle
\delta \rho_h(k)^2 \rangle/(2 \pi^2)$, where $\delta \rho_h(k)$ is the
fluctuation in the halo mass density in Fourier space.  The
$\Delta_{hh}^2$ of the merger history tree plus resolved halos ({\it
solid curve}) agrees to better than $40 \%$ at all scales with the
small box $\Delta_{hh}^2$ ({\it dotted curve}).\footnote{Because the
$20 \,h^{-1}$ Mpc box is missing modes larger than the size of the
box, we expect that it underestimates the true spectrum of
cosmological fluctuations \citep{barkana03}. A larger box with the
same mass resolution would result in better agreement between the
solid and dotted curves.} The level of agreement between the dotted
and solid curves demonstrates that the merger tree method reliably
incorporates the small mass halos in our simulations. The thin solid
curve is an analytic prediction for the halo power spectrum given by
$k^3 \, \bar{b}_{PS}^2 \, P_{\delta \delta}/(2 \pi^2)$, in which
$P_{\delta \delta}$ is calculated using the Peacock and Dodds fitting
formula for the density power spectrum, $\bar{b}_{PS} = \int_{m_{\rm
cool}}^\infty dm \, m \, n_{PS}(m) \, b_{PS}(m)$, and $ b_{PS}(m)$ is
the Press-Schechter bias for a halo of mass $m$ [at $z = 8.7$,
$\bar{b}_{PS} = 3.5$] \citep{mo96}.  This analytic estimate for
$\Delta_{hh}^2$ ignores Poisson fluctuations, and a comparison with
the other curves indicates that Poisson fluctuations are important on
scales of $k \gtrsim 4 \, h \; \Mpc^{-1}$.

The top panel in Figure \ref{fig:mtps} shows the mass-weighted power
spectrum of halos above the cooling threshold from the merger history
tree method ({\it solid curves}) and of the halos that are well resolved in
our box with $m > 2 \times 10^9 ~ M_{\odot}$ ({\it dashed curves}) at $z
=6.6$ ({\it thin curves}) and $z = 11.1$ ({\it thick curves}).  The different
spectrum of fluctuations between the solid and dashed curves
suggests that incorporating the unresolved halos may lead to a different
HII morphology.    As the source halos become
rarer, their spatial fluctuations increase and the Poisson component
of the fluctuations becomes more important.

\begin{table*}
\begin{center}
\caption{Radiative transfer simulations discussed in this
paper. Unless otherwise specified, the subgrid clumping factor $C_{\rm
cell}$ is set to unity and the radiative transfer is performed on a
$256^3$ grid.  $M_8$ denotes the halo mass in units of $10^8
~M_{\odot}$.  The functions $C_{S2}, ~C_{S3}$ and $C_{S4}$ are
calibrated such that the sources in the respective simulations output
the same number of ionizing photons in each time step as the sources
in simulation $S1$.  }

\end{center}
\begin{center}
\begin{tabular}{l c c l}
\hline
Simulation & {Merger Tree Halos$^*$} & {$\dot{N}$ (photons s$^{-1}$)} &
{Comments}\\

\hline
 S1 & yes & $2 \times 10^{49} \, M_8 $ & \\
 S2 & yes & $C_{S2} \, M_8^{1/3}$ & \\
 S3 & yes & $C_{S3} \, M_8^{5/3}$ & \\
 S4 & no & $C_{S4} \, M_8$ & includes only $m > 4 \times
 10^{10} ~ M_{\odot}$ \\
F1 & yes & $2\times10^{49} \, M_8$ & feedback on $m < M_{\rm J}/2$; $\tau_{SF} = 100$ Myr \\
F2 & yes & $2\times10^{49} \, M_8$ & feedback on $m < M_{\rm J}/2$; $\tau_{SF} = 20$ Myr \\
F3 & no & $2\times10^{49} \, M_8$ & includes only halos with $m > M_{\rm J}/2$ \\
 C1  & no & $3 \times10^{49} \, M_8$ & all cells set to mean density\\
 C2  & no & $3 \times10^{49} \, M_8$ &\\
 C3 & no & $3 \times 10^{49} \, M_8 $ & $512^3$ grid\\
 C4  & no &$6 \times10^{49} \, M_8$ & $C_{\rm cell}$ given by eqn. \ref{eqn:Ccell}\\
 C5  & no &$6 \times10^{49} \, M_8$ & $C_{\rm cell} = 4 + 3 \, \delta_{\rm cell}$ \\
M1 & no & $9 \times 10^{49} \, M_8$ & \\
M2 & no & $9 \times 10^{49} \, M_8$ &  \citet{iliev-mh} minihalos with $m_{\rm mini} > 10^5 M_{\odot}$\\
M3 & no  & $9 \times 10^{49} \, M_8$ & minihalos with $m_{\rm mini} > 10^5 M_{\odot}$, $\sigma_{\rm mh} = \pi r_{\rm vir}^2$\\
Z1 & yes & $1 \times 10^{50} \, M_8$ & \\
Z3 & yes & $5 \,C_{S3} \, M_8^{5/3}$ & \\
\hline
\end{tabular}
\end{center}  
$^*$ All radiative transfer simulations are post-processed on a
density field that resolves halos down to $10^9 \, M_{\odot}$.  Halo
mass resolution is extended beyond $10^9 \, M_{\odot}$ with a merger
tree.  Here, `yes' means the source halo
resolution is supplemented with the merger tree down to $m_{\rm
cool}$.
 \label{table1}
\end{table*}

\section{Sources}
\label{sources} 
Now that we have a method for incorporating small mass halos into our
simulations, we examine several prescriptions for populating the dark
matter halos with ionizing sources.  We consider models where
POPII-like sources are responsible for the vast majority of the
ionizing photons.  Even among these sources, it is uncertain which
galaxies will produce the ionizing photons.  We consider four models
for the source efficiencies.  In all models, the ionizing luminosity
$\dot{N}$ for a halo of mass $m$ is given by the relation $\dot{N}(m)
= \alpha(m) \, m$.  In simulation S1, the factor $\alpha$ is
independent of halo mass. Simulation S2 uses the same source halos as
S1 except $\alpha \propto m^{-2/3}$ (the lowest mass systems are the
most efficient at converting gas into IGM ionizing photons). In
simulation S3, we again use the same source halos but set $\alpha
\propto m^{2/3}$ (the most massive systems are the most
efficient). Finally, in simulation S4, $\alpha \propto m^0$, as in S1,
except that only halos with $m > 4\times10^{10} ~M_{\odot}$ are
sources. At $z=9$, there are $500$ sources in S4 and, at $z=7$, there
are $7000$ sources. These numbers are in contrast to the other
simulations in this section in which there are over $1$ million
sources at $z = 9$ and over $3$ million at $z = 7$.

Table 1 lists the parameters we use for the runs in this section.  For
simulation S1 we set $\dot{N}(m) = 2\times10^{49}\, m/(10^8 \,
M_{\odot})$ photons s$^{-1}$.  To facilitate comparison, we normalized
the photon production in the S2, S3 and S4 runs so that the same
number of photons are outputted in each time step as
in S1.  In reality, as rarer sources dominate the ionizing budget, the
rate at which the Universe is ionized quickens because the number of
high mass halos is growing exponentially. Here we are interested in
the structure of reionization, which is not
significantly affected by the duration of this epoch.

The luminosity of our sources only depends on the
halo mass.  This parametrization is most reasonable if, once the gas has
cooled within a halo, the timescale for its conversion into stars is
at least comparable to the duration of reionization (or a few hundred million
years). \citet{springel03} measure a gas-to-star conversion timescale
of over a gigayear in simulations of high redshift galaxies.  However,
many works in the literature parameterize star formation as
proportional to the time derivative of the collapse fraction
(e.g., \citet{furlanetto04a}). This parametrization assumes that the
rate at which a galaxy converts its cold gas into stars is much shorter
than the duration of reionization. The effects of alternative
parameterizations of star formation on reionization are discussed at the
end of this section.

The source prescriptions in S1, S2, S3 and S4 are all still
reasonable in principle. The least massive systems could dominate the
budget of ionizing photons because it may be easier for ionizing
photons to escape from the smallest mass halos. \citet{wood99} find
that this is the case in static halos owing to the shallower potential well
of the low mass halos.  Internal feedback from galactic winds and
supernova bubbles may further enhance the escaping luminosity of
smaller halos relative to the more massive halos. Internal feedback
can also act to shut off star formation.  \citet{springel03} find that
feedback from galactic winds suppresses star formation in the least
massive systems relative to the more massive. The scaling
$\alpha \sim m^{2/3}$ taken in model S3 is motivated by the observed star
formation efficiency in low redshift dwarf galaxies
\citep{kauffmann03}.

Because star formation is a complicated process, observations rather
than theory will likely drive our knowledge of the high redshift
sources. From present observational constraints, the source
prescription used in S4 is closest to being ruled out:  There is
mounting evidence that the highest mass halos cannot
produce enough photons to ionize the Universe \citep{stark06}.

All the simulations in this section were performed on a $256^3$ grid,
and the subgrid clumping factor is set to unity (i.e., density
fluctuations on scales smaller than the cell scale are ignored). In
subsequent sections, we increase the level of clumping and include
dense absorbing systems that limit the mean free path of photons.
Due to the lack of gas clumping in the runs in this section, our
simulations underestimate the number of ionizing photons needed to
reionize the IGM.  However, we find that neither the dense absorbers
nor the increased clumping have a substantial effect on the topology
of the bubbles for fixed $\bar{x}_i$, except in extreme scenarios or
at higher ionization fractions than we consider.

\begin{figure*}
\begin{center}
{ \epsfig{file=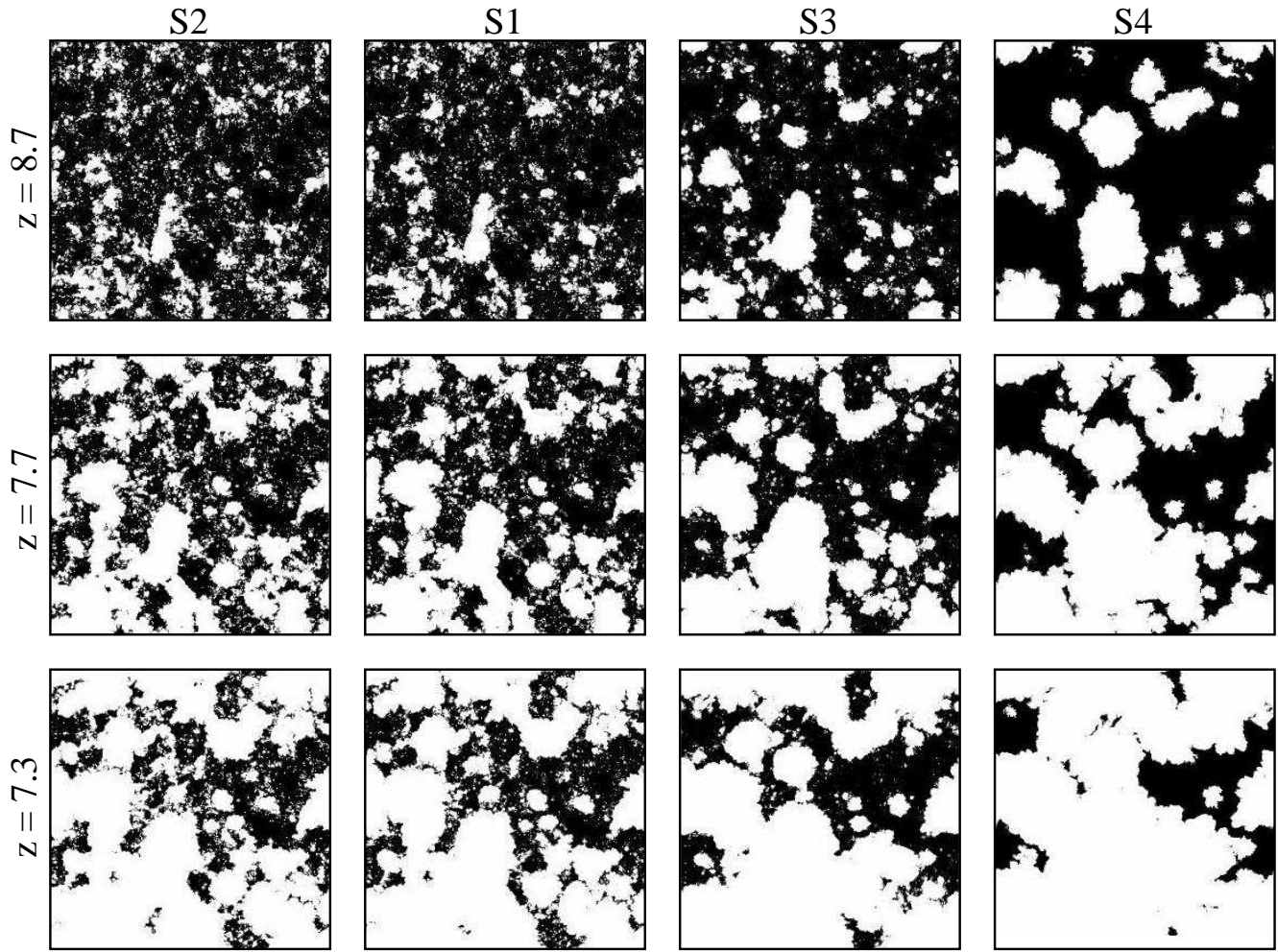, width=17.6cm}}
\end{center}
 \caption{Comparison of four radiative transfer simulations
post-processed on the same density field, but using different source
prescriptions parameterized by $\dot{N}(m) = \alpha(m) \, m$.  The
white regions are ionized and the black are neutral.  The left,
left-center, right-center and right panels are respectively cuts
through simulations S2 ($\alpha \propto m^{-2/3}$), S1 ($\alpha
\propto m^0$), S3 ($\alpha \propto m^{2/3}$), and S4 ($\alpha \propto
m^0$, but only halos with $m > 4\times 10^{10} \, M_{\odot}$ host
sources).  For the top panels, the volume ionized fraction is
$\bar{x}_{i, V} \approx 0.2$ (the mass ionized fraction is
$\bar{x}_{i, M} \approx 0.3$) and $z = 8.7$. For the middle panels,
$\bar{x}_{i, V} \approx 0.5$ ($x_{i, M} \approx 0.6$) and $z = 7.7$,
and for the bottom panels, $\bar{x}_{i, V} \approx 0.7$ ($\bar{x}_{i,
M} \approx 0.8$) and $z = 7.3$. Note that the S4 simulation outputs
have the same $\bar{x}_{i, M}$, but $\bar{x}_{i, V}$ that are typically
$0.1$ smaller than that of other runs.  In S4 the source fluctuations are
nearly Poissonian, resulting in the bubbles being uncorrelated with
the density field ($\bar{x}_{i, V} \approx \bar{x}_{i, M}$). Each
panel is $94 ~\Mpc$ wide and would subtend $0.6$ degrees on the
sky.\label{fig:sources}}
\end{figure*}

Figure \ref{fig:sources} compares slices through the ionization field
from the S2, S1, S3, and S4 simulations ({\it left to right}) at
redshifts $8.7, 7.7$ and $7.3$ ({\it top, middle and bottom panels}).
Panels in a row have the same mass ionized fraction $\bar{x}_{i,M}$.
All panels have bubbles located around the large-scale overdense
regions, but the bubbles better trace the
overdensities as the less massive sources dominate.  Reionization in
both S1 and S2 is dominated by the low mass sources and results in a
nearly identical reionization morphology when comparing at fixed
$\bar{x}_i$.  The HII regions in S3 are larger and more spherical than
they are in S1 and S2.  The bubbles are still larger in S4.

\begin{figure}
\rotatebox{-90} { \epsfig{file=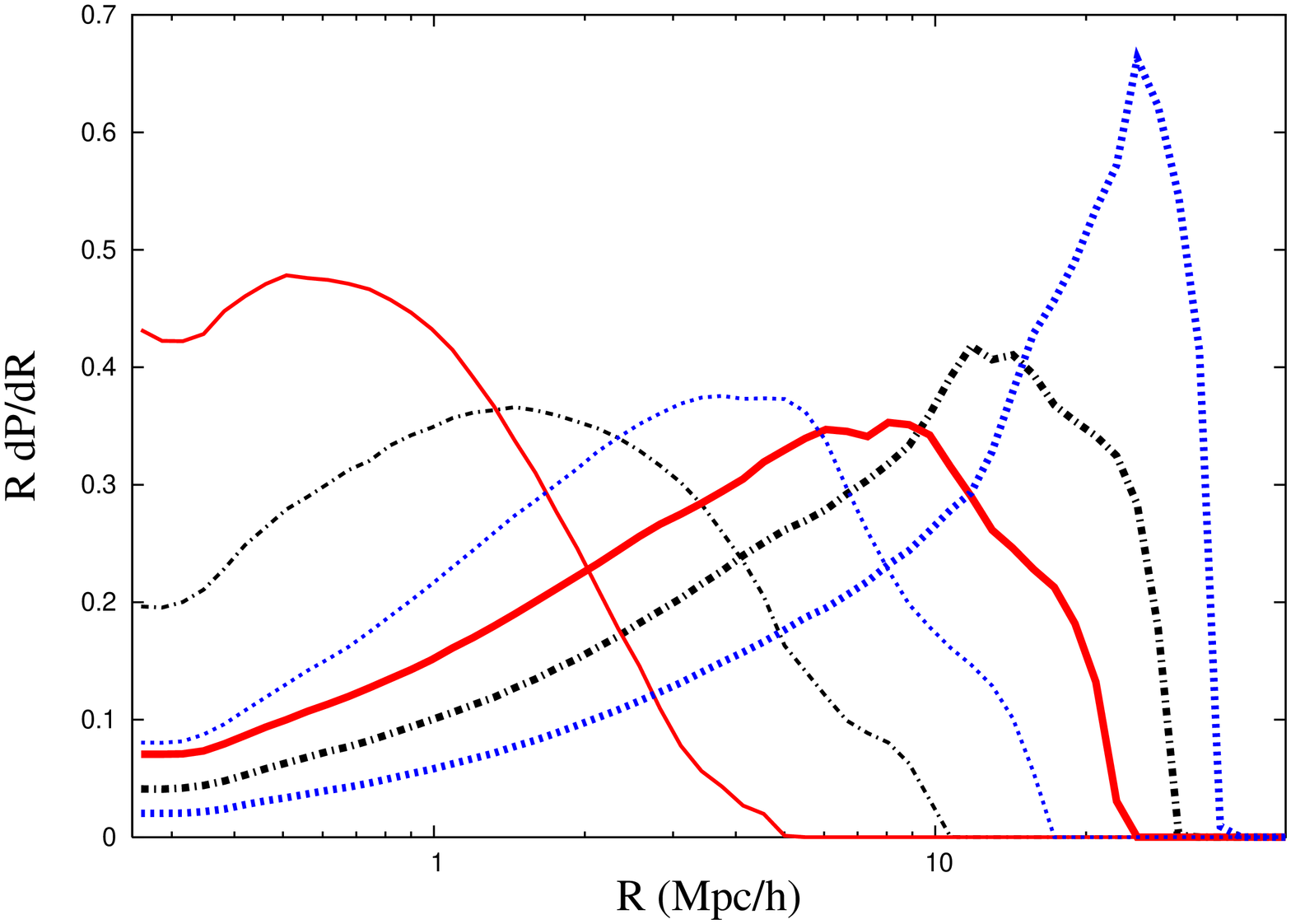, width=7.cm, height = 8.7cm}}
 \caption{ The volume-weighted bubble radius PDF for the S1 ({\it
solid curves}), S3 ({\it dot-dashed curves}), and S4 ({\it dotted
curves}) simulations.  See the text for our definition of the bubble
radius $R$. We do not include curves for the S2 simulation because
they are similar to those for S1.  The thin curves are at $z = 8.7$
and $\bar{x}_{i,M} = 0.3$, and the thick curves are at $z = 7.3$ and
$\bar{x}_{i, M} = 0.8$.  Simulation S4 has the rarest sources and the
largest HII regions of the four models. \label{fig:sourcesbub}}
\end{figure}

\begin{figure}
\rotatebox{-90}{
 \epsfig{file=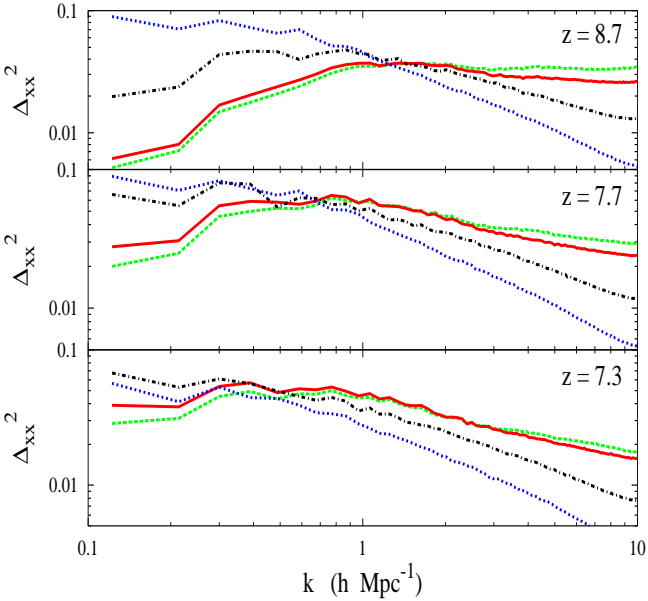, width=8cm, height=8.8cm}}
 \caption{The ionization fraction power spectrum $\Delta_{xx}(k)^2 =
k^3 P_{xx}(k)/2 \pi^2$ for the S1 ({\it solid curves}), S2 ({\it
dashed curves}), S3 ({\it dot-dashed curves}) and S4 ({\it dotted
curves}) simulations.  For the top panels, $\bar{x}_{i, V} \approx
0.2$ ($\bar{x}_{i, M} \approx 0.3$), for the middle panels,
$\bar{x}_{i, V} \approx 0.5$ ($x_{i, M} \approx 0.6$), and for the
bottom panels, $\bar{x}_{i, V} \approx 0.7$ ($\bar{x}_{i, M} \approx
0.8$). In all panels, the fluctuations are larger at $k \lesssim 1 ~ h
\, \Mpc^{-1}$ in S3 and S4 than they are in S1 and S2.  As the most
massive halos contribute more of the ionizing photons, the ionization
fraction fluctuations increase at large scales.\label{fig:pksources}}
\end{figure}

The differences between the ionization maps owe to the bias
differences between the sources.  As the sources become more biased,
they become more clustered around the densest regions, resulting in
the bubbles becoming larger.  In Press-Schechter theory, the
luminosity-weighted average bias at $z=8$ is $\bar{b}_{\rm PS} = 2.8$
for the S2 source prescription, $3.2$ for the S1, $5.0$ for the S3,
and $7.3$ for the S4.  The S4 sources are located in the highest
density peaks in the Universe and the fluctuations in the density of
these sources is the largest.  [{\it See respectively, the thin solid,
thin dotted, and thin dash-dotted lines in Fig. \ref{fig:mtps} for a
comparison of the luminosity-weighted power spectra for the S1, S3,
and S4 sources at $z = 6.6$.}]  The differences between the ionization
maps for S1, S3, and S4 should allow observations to
distinguish between these scenarios (as discussed in \S
\ref{observations}).

This trend of bubble size increasing with average source mass was
predicted in the analytic work of \citet{furl-models}.  Analytic
models typically ignore Poisson fluctuations in the source abundance,
which can dominate over cosmological fluctuations when relatively
massive sources dominate the photon production.\footnote{While it is
difficult to incorporate Poisson fluctuations in analytic models based
on the excursion set formalism, \citet{furl-models} investigated the
effect of Poisson fluctuations using such models.}  For example, the
bubble scale in S4 is roughly $20$ Mpc at $z = 7.7$ and $\bar{x}_i =
0.5$ -- a scale where Poisson fluctuations dominate over the
cosmological ones.  For the S1 and S2 source models, cosmological
fluctuations dominate over Poisson fluctuations on the scale of a
typical bubble, but Poisson fluctuations can be important in smaller
bubbles.  This deficiency of analytic models was noted in
\citet{zahn06}.

Figure \ref{fig:sourcesbub} plots the bubble size distribution from S1
({\it solid curves}), S3 ({\it dot-dashed curves}), and S4 ({\it
dotted curves}) at $\bar{x}_{i, M} = 0.3$ ({\it thin curves}) and
$\bar{x}_{i, M} = 0.8$ ({\it thick curves}). The S2 simulation is not
included here; it yields bubble sizes that are similar to those in S1.
How do we define the bubble ``radius'' since the bubbles are far from
spherical?  For each cell in the box, we find the largest sphere
centered around this cell that is $90\%$ ionized.  We say that each
cell is in a bubble of size equal to the radius of this sphere.  We
then tabulate the radius from all the ionized cells to calculate the
volume-weighted bubble PDF (zero-radius bubbles are not included in
the tabulation).  This definition of bubble size is chosen to
facilitate comparison with analytic models of reionization based on
the excursion set formalism in which the bubble radius is similarly
defined \citep{furlanetto04a}. The bubbles are largest in S4 and
smallest in S1, and in all runs there is a characteristic bubble
radius.

It is useful to compare the measured bubble size distribution to the
size distribution predicted in analytic models. The ``log-normal''
distribution of bubbles found in analytic models is present in these
simulations.  The bubble size distribution becomes more sharply peaked
in $\log(R)$ with increasing $\bar{x}_i$ in our simulations, a trend
that was predicted by analytic models \citep{furl-models}.  A more
detailed comparison of the bubble sizes between these simulations and
analytic models is given in \citet{zahn06}.

Figure \ref{fig:pksources} plots the dimensionless ionization fraction
power spectrum $\Delta_{xx}^2$ for the four simulations ({\it S1 --
solid, S2 -- dashed, S3 -- dot--dashed, S4 -- dotted}) for the volume
ionized fractions $\bar{x}_{i,V} \approx 0.2$ ({\it top panel}),
$\bar{x}_{i,V} \approx 0.5$ ({\it middle panel}) and $\bar{x}_{i,V}
\approx 0.7$ ({\it bottom panel}). [The $\bar{x}_{i,V}$ for S4 is
$\approx0.1$ smaller than these values.]  Note that $\Delta_{xx}(k)^2
= k^3 \, P_{xx}(k)/(2 \pi^2)$ in which $(2 \pi)^3 \, P_{xx}(k)\,
\delta_D(\bfk - \bfk') = \langle x_i(\bfk) x_i(\bfk') \rangle $.  For
some $\Delta_{xx}^2$ the power peaks at the box scale ($k \approx
0.1~\Mpc^{-1}$), particularly at larger $\bar{x}_i$.  This indicates that
there is substantial power in ionization fraction fluctuations on
scales larger than our simulation box in some of the considered
models.  Therefore, the box we use is too small to make statistical
predictions about reionization for some of the models and at some
$\bar{x}_i$.  Lyman-limit systems or minihalos may reduce the size of
the largest bubbles and alleviate this difficulty (see \S
\ref{minihalos}).

It is useful to note that an ionization field that is composed of
fully neutral and ionized regions with total ionized fraction
$\bar{x}_{i, V}$ has variance of $\bar{x}_{i,V} - \bar{x}_{i,V}^2$
on small scales, implying that
\begin{equation}
\int_0^\infty d\log k ~\Delta_{xx}(k)^2 = \bar{x}_{i,V} -
\bar{x}_{i,V}^2. \label{eqn:xxpower}
\end{equation}
Because of equation (\ref{eqn:xxpower}) and because the snapshot from
S4 has more power on large scales, the snapshots from S1, S2, and S3
must have more power than S4 on small scales for the same $\bar{x}_i$. The
distribution of power has important implications for upcoming
observations.  Generally speaking, the more power on large scales ($k <
1 \,h \; \Mpc^{-1}$), the more observable the signal (see \S
\ref{observations}).

The picture of reionization seen in simulations S1, S2 and S3 is
different from that seen in the simulations of \citet{iliev05}.  Their
simulations resolve halos with $m > 2\times 10^9 \; M_{\odot}$, and
reionization ends at $z \approx 12$ in their calculations.  Hence, the
resolved halos in their simulations are very rare and, of the four
source models we consider, are most similar in abundance to the source
halos in S4. Their reionization snapshots give the visual impression
of many overlapping spheres. We do see, particularly in simulation S4,
that the bubbles become more spherical as the sources become rarer.
See \citet{zahn06} for further comparison.

The prescription we use for the luminosity of the sources is
simplistic.  In all of our source models, the luminosity of a halo is
monotonic in the halo mass such that the characteristic source mass is
either $m_{\rm cool}$ or $m_*$ -- the halo mass that characterizes the
transition to the exponential tail in the luminosity function.  Star
formation is complicated, and the characteristic mass of a source could
be an intermediate mass between $m_{\rm cool}$ and $m_*$.  In this case, the
bias of the sources will fall between the source bias in S2 and in
S3, and, therefore, the bubbles sizes will be between the sizes in S2
and in S3 if we compare at fixed $\bar{x}_i$.

Surely the luminosity of galaxies depends on additional parameters
besides the halo mass. Other studies have parameterized the luminosity
of the sources as proportional to the time derivative of the collapse
fraction, considering the accretion of gas onto sources as a better
proxy for the star formation rate than the gas mass of the sources.
We have run simulations with the luminosity proportional to the time
derivative of the collapse fraction in a cell.  We find that the
morphology of reionization is very similar between this
parameterization and that of the constant mass-to-light model. The
reason for this similarity is that the collapse fraction in a given
region is changing nearly exponentially with time and so the rate of
halo mass growth is proportional to the halo mass.  Alternatively,
star formation or quasar activity may be correlated with major mergers
(see Hopkins et al. 2006a,b, Li et al. 2006 for discussion).  Since
major merger events are more biased, this results in larger
bubbles. \citet{cohn06} used an analytic model to derive the bubble
sizes in merger-driven scenarios.  In addition, it might have been
possible for the gas in smaller mass galaxies ($m \gtrsim 10^5
~M_{\odot}$) to cool via $H_2$ transitions. If this is the case, stars
would form in halos with smaller masses than are considered here.
These sources would be less biased, and, therefore, the HII regions
would be smaller and more fragmented.\footnote{If molecular hydrogen
cooling does happen at low redshifts, then it may occur in
halos with $m \sim 10^7~ M_{\odot}$.  Feedback processes may
destroy the $H_2$ in smaller halos.  However, $\bar{b}_{\rm PS} = 2.6$
for halos with $m > 10^7 ~M_{\odot}$ at $z = 8$, as opposed to
$\bar{b}_{\rm PS} = 2.8$ in S2, such that the bubble sizes will be
similar to the sizes in S2.  The harder spectrum of POPIII stars will
make the ionization fronts less sharp.}

\section{Source Suppression from Photo-heating} \label{feedback}

The extent to which photo-heating from a passing ionization front
affects the ionizing sources and, as a result, the reionization
process is not well understood. Often, when included in a study, the
effect of photo-heating is parameterized in a simplistic fashion: Star
formation is assumed to be completely shut off in the low mass sources
as soon as an ionizing front has passed.  However, sources that form
prior to a front passing will have a cool reservoir of gas with which
to make stars.  Since photo-heating can suppress subsequent accretion
onto these objects, eventually this reservoir will run dry and all the
gas will have been converted to stars.  The timescale over which this
reservoir will be depleted is uncertain (see discussion in \S
\ref{sources}).

Furthermore, the mass threshold at which sources will be suppressed by
photo-heating is fairly unconstrained.  Often the suppression mass
scale is taken to be the linear theory Jeans mass $M_{\rm J}$.  This
choice is, however, problematic.  The gas will not instantaneously
respond to photo-heating -- there will be some delay, leading to a
time dependent suppression threshold that only asymptotically
approaches the Jeans mass for linear fluctuations \citep{gnedin98}. In
addition, a spherical perturbation that collapses at $z = 8.0$ was at
turnaround at $z=13.3$. An ionization front passing this collapsing
mass at, say, $z=9$, will do little to prevent the gas from cooling.
The collapsing gas is already significantly overdense prior to
front-crossing, giving it a large collisional cooling rate and
possibly allowing it to self shield \citep{dijkstra04}.
\citet{dijkstra04} finds in 1-D simulations that a substantial
fraction of collapsing density peaks with mass below the Jeans mass
threshold (or $2.7\times 10^9~ M_{\odot}$ at $z = 7$ for $T_{\rm gas}
= 10^4$ K) are still able to collapse and form gas-rich halos in
ionized regions, and \citet{kitayama00} and \citet{kitayama01} find an
even larger fraction than \citet{dijkstra04} in 3-D simulations.

\citet{iliev06b} was the first to investigate with large-scale
simulations of reionization the effect feedback on the sources from
photo-heating has on reionization.  They applied the rather extreme
criterion that star formation in all halos below $10^9 ~M_{\odot}$ is
shut off after $20$ million years in ionized regions. They concluded
from this study that the small halos do not play an important role in
ionizing the IGM.  Here we expand upon the work of \citet{iliev06b} to
include more general parameterizations for the feedback from
photo-heating. 

\begin{figure}
\rotatebox{-90}{\epsfig{file=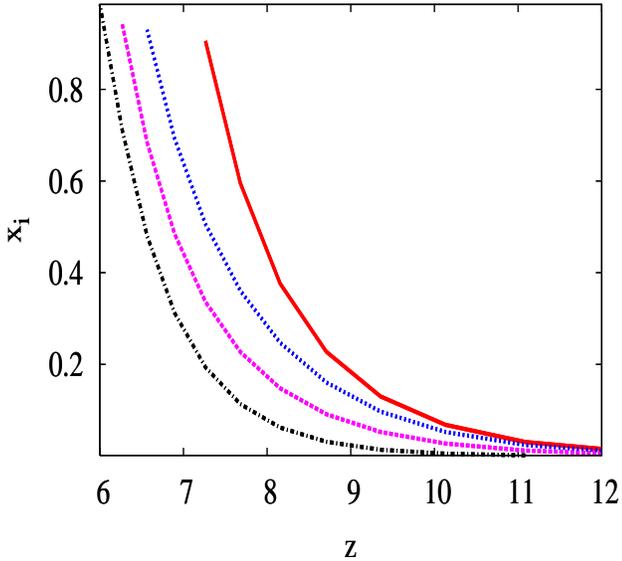, width=8cm, height=8.8cm}}
\caption{The impact of thermal feedback on the ionization history.  The
ionization history $\bar{x}_{i,V}(z)$ for the S1 ({\it solid curve}), F1
({\it dotted curve}), F2 ({\it dashed curve}), and F3 ({\it dot-dashed
curve}) simulations.  Simulation S1 has no feedback, and F3 has maximal
feedback.  Feedback extends the duration of reionization by as much as
$200$ Myr in our simulations.}
\label{fig:feedback_xi}
\end{figure}

\begin{figure}
\begin{center}
 \epsfig{file=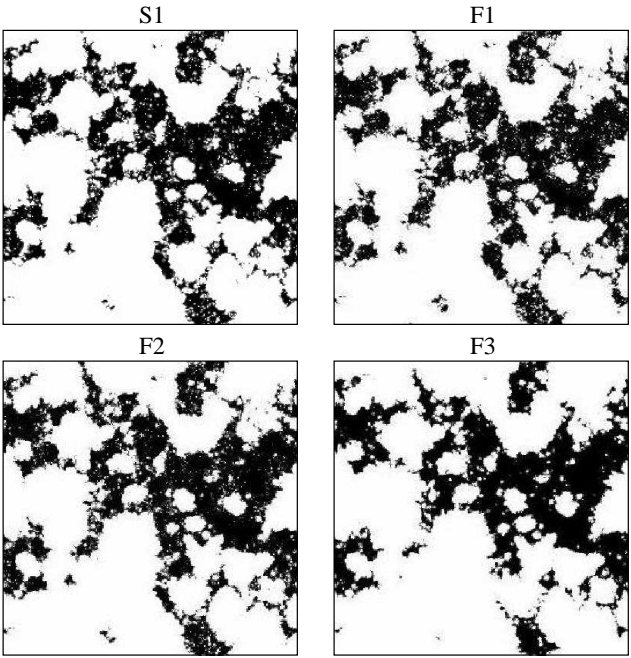, width=8.4cm}
 \end{center}
\caption{The impact of thermal feedback on the morphology of
reionization.  Slices through snapshots in which $\bar{x}_{i, V}
\approx 0.7$.  The white regions are ionized and the black are
neutral.  S1 includes no thermal feedback, F1 has minimal feedback
with $\tau_{SF} = 100$ Myr, F2 has strong feedback with
$\tau_{SF} = 20$ Myr, and F3 uses only halos with $m >
M_{\rm J}/2$ as the sources (or, equivalently, $\tau_{SF} =
0$). Notice that even the ionization maps from S1 and F3 do not
differ by much, which shows that feedback has a small effect on the
structure of reionization.}\label{fig:feedback}
\end{figure}

The parameterizations we adopt for source suppression owing to
photo-heating are simplistic.  However, we show that the structure of
reionization is largely unaffected by feedback even for an aggressive
parametrization of suppression.  If an ionizing front passes a source
with luminosity $L_0$ at time $t_i$ then at time $t$ we set its
luminosity to be $L(t) = L_0 \, \exp[-(t-t_i)/\tau_{SF}]$, where
$\tau_{SF}$ can be thought of as the timescale over which the cool gas in
the potential well of a source is converted into stars.  We set
$\tau_{SF} = 100, ~ 20$ and $0$ million years in simulations F1, F2,
and F3, respectively.  We assume that this luminosity suppression
affects halos with masses below $M_{\rm J}/2$, where $M_{\rm J}$ is
calculated for gas at $T = 10^4 \, K$ (or $1.4 \times 10^9 ~M_{\odot}$
at $z = 7$).  This fixed suppression mass misses the time dependent
response of the gas to photo-heating. The suppression mass $M_{\rm
J}/2$ is approximately the suppression mass found at $z = 7$ in
\citet{dijkstra04}.  This suppression mass is an order of magnitude
larger than that found by \citet{kitayama00}. Furthermore, we assume
that halos that form in already ionized regions with masses below
$M_{\rm J}/2$ have zero star formation and do not contribute to
reionization.\footnote{For simplicity, we take the sources that exist
with masses below $M_{\rm J}/2$ at the instant a region becomes
ionized to be the sources with $m < M_{\rm J}/2$ that contribute
photons for all subsequent times.  In reality, a fraction of these
halos that form prior to front crossing will become incorporated in
more massive halos than $M_{\rm J}/2$, halos we count as separate sources.
Therefore, we double count some of the mass in halos and underestimate
the effect of feedback when $\tau_{\rm SF} >0$.  This underestimate
does not change our conclusions.}

First, in agreement with previous studies, we note that thermal
feedback can delay and extend the reionization process (Fig.
\ref{fig:feedback_xi}).  Simulation S1 ({\it solid curve}) does not include
feedback, whereas simulation F3 ({\it dot-dashed}) includes maximal feedback
($\tau_{SF} = 0$).  The duration of reionization is extended by about
$200$ million years in this case.  For the other feedback scenarios
(F1 and F2), reionization is extended by a shorter period ($100$ and
$150$ million years).

Figure \ref{fig:feedback} displays slices through snapshots with
$\bar{x}_{i,V} = 0.7$ for the simulations S1, F1, F2, and F3.  In S1,
halos below the $m_{\rm cool}$ always contribute ionizing photons,
whereas in simulation F3 only halos above $M_{\rm J}/2$
contribute. The differences between S1 and F3 are minor: The small
mass sources do not change the structure of reionization
significantly.  S1 has more small bubbles, and the HII fronts have more
small-scale features. Simulation F1 ($\tau_{SF} = 100$ Myr) is most
similar to S1 -- long gas-to-star formation timescales essentially
negate the effect of feedback, and simulation F2 ($\tau_{SF} = 20$
Myr) has less structure in the voids than F1.  In conclusion, simulations S1 and F1--3 have a
very similar morphology at fixed $\bar{x}_i$.  Feedback does not significantly affect the
structure of reionization.  We find that this conclusion still holds
if we compare at other $\bar{x}_{i,V}$ as well.

To make the comparison of feedback models more quantitative, we
contrast the $\Delta_{xx}$ at $\bar{x}_i = 0.7$ for these four models.
We find that the $\Delta_{xx}$ of the S1, F1, and F2 models agree to
approximately $10\%$ at all scales and that the $\Delta_{xx}$ of the S1
and F3 models (no feedback and maximal feedback models) differ by at
most $20\%$, with the largest differences being for modes near the box
scale and for modes with $k > 5 \, h \; \Mpc^{-1}$.

It is simple to understand why thermal feedback has little impact on
the size distribution and morphology of HII regions (provided we
compare at fixed $\bar{x}_i$).  The bubble size distribution and
morphology are mainly sensitive to the bias of the ionizing source
host halos and to Poisson fluctuations in the halo abundance for
sufficiently rare source halos.  The top panel in Figure \ref{fig:mtps}
compares the luminosity-weighted halo power spectrum $\Delta_{hh}^2$
for halos above the cooling mass at $z =6.6$ ({\it thin, solid curve})
compared to $\Delta_{hh}^2$ for halos with $m > 2\times 10^9 ~
M_{\odot}$ ({\it thin, dashed curve}).  Notice that the difference
between these curves is less than the difference between, for example,
these curves and those for the S3 sources ({\it thin, dotted curve}).
In terms of the Press-Schechter bias at $z=8$, $\bar{b}_{\rm PS} =
3.2$ for the S1 sources whereas $\bar{b}_{\rm PS} = 4.3$ for halos
with $m > M_{\rm J}/2 =  1.4 \times 10^9 ~ M_{\odot}$.  These values should be
contrasted with $\bar{b}_{\rm PS} = 5.0$ for the S3 sources and
$\bar{b}_{\rm PS} = 7.3$ for the S4 sources.  Therefore, if halos with
$m \lesssim 1.4 \times 10^9 ~M_{\odot}$ are evaporated (as in this
section), the morphology of reionization is not changed as
substantially as the difference between the morphology in the S1 and
in the S3/S4 simulations.  In fact, Figure \ref{fig:feedback} shows
that the bubbles are largely unchanged by feedback.

All simulations in this section are parameterized such that $\dot{N}
\propto m$ and such that the suppression scale is $M_{\rm J}/2$.  For
lower $\bar{x}_{i,V}$ than are shown in Figure \ref{fig:feedback}, the
effect of feedback in our simulations is even less significant. If the
highest mass sources are more efficient at producing ionizing photons,
reionization will be extended by a smaller amount by feedback than we
find, whereas if the low mass sources are more efficient, feedback will extend reionization by a larger amount.  The conclusion that the structure
of reionization is only modestly affected by feedback holds even if
the sources near $m_{\rm cool}$ are more efficient at producing
ionizing photons then we have assumed: We found in \S \ref{sources}
that as we made the low mass sources more efficient, the properties of
the HII regions are essentially unchanged (compare the panels from S1
and S2 in Fig. \ref{fig:sources}). Lastly, we believe that our choice
of $M_{\rm J}/2$ is a fairly extreme suppression mass for low
redshift, POPII star reionization scenarios owing to effects mentioned
at the beginning of this section.  If the suppression mass is larger
than $M_{\rm J}/2$ or if reionization happens at a higher redshift but
with the same suppression mass, thermal feedback will be more
important. However, at $z>10$ both \citet{dijkstra04} and
\citet{kitayama00} find that the suppression mass is much lower than
$10^9 ~M_{\odot}$ .

If molecular hydrogen cooling is able to cool the gas in a halo to
form a galaxy then most star formation could take place in halos with
$m \ll  m_{\rm cool}$.  In such a case, thermal feedback could play a
more important role in shaping the structure of reionization.
\citet{kramer06} found that this scenario could lead to a bimodal
bubble size distribution.  (Note that in the models that we consider
in which only halos with $m > m_{\rm cool}$ form stars, feedback does
not create a bimodal bubble size distribution, and the size
distribution of the bubbles is largely unchanged by thermal feedback.)

\section{Effect of Density Inhomogeneities}
\label{recombinations}

Density inhomogeneities potentially play an important role in shaping
the HII regions during reionization.  On small scales, density
inhomogeneities lead to the outside-in reionization observed in the
simulations of \citet{gnedin00}.  The role of these inhomogeneities on
the large-scale bubble morphology has not been investigated in
detailed simulations.  Analytic models make simplistic assumptions to
incorporate their effects. These models spherically average the
density fluctuations in a bubble and typically treat a higher level of
recombinations as equivalent to decreasing the ionizing efficiency of
the sources.

Previous large-scale radiative transfer simulations of reionization
either ignored subgrid density inhomogeneities entirely, or they
calibrated their subgrid clumping factor from smaller simulations
\citep{mellema06, kohler05}. A simulation of \citet{mellema06} uses a
clumping factor that is independent of $\delta$ and $\bar{x}_i$ and
neither the simulations of \citet{mellema06} nor \citet{kohler05}
include a dispersion in the clumping for a cell of a given
overdensity. Both studies of clumping also assume that the clumping
factor is independent of the local reheating and ionization history,
which is incorrect in detail.  In linear theory, the smallest gas
clump -- which is intimately tied to the gas clumping factor -- is
given by the filtering mass $M_f$ \citep{gnedin98}, and this mass
incorporates the time-dependent gas response to heating (see \S
\ref{gnedin}).  The filtering mass provides some framework to
understand the small-scale gas clumping.  It is important to
understand how sensitive the characteristics of reionization are to
gas clumping -- to what extent can gas clumping be ignored or included
in only a primitive manner?

Minihalos -- virialized objects with $T_{\rm vir} < 10^4$ K --
contribute to the clumping differently than does the diffuse IGM.
These virialized objects are unresolved in all current large-scale
reionization simulations.  Minihalos are extremely dense and act as
opaque absorbers until they are photo-evaporated.  Since the inner
regions of minihalos are self-shielded, it is difficult to describe
the effect of minihalos with a subgrid clumping factor.  In addition,
most photons that pass through a cell should not be affected by a
minihalo because the mean free path for a ray to intersect a minihalo
can range between $1$ and $100$ Mpc.  Absorptions by minihalos are
unimportant when the HII regions are much smaller than the mean free
path.  Once the bubble size becomes comparable to the mean free path,
minihalos may be the dominant sinks of ionizing photons within a
bubble.  \citet{furlanetto05} predict that minihalos create a sharp
large-scale cutoff in the size distribution of bubbles, particularly
when the Universe is largely ionized. If this prediction is true,
large scale topological features during reionization can be used to
probe small-scale density fluctuations.

We split the discussion in this section into two components: (1)
quasi-linear IGM density inhomogeneities, and (2) the minihalos. (Our
discussion on the effect of minihalos also applies to the effect of a
more general class of dense absorbers, Lyman-limit systems.)  The
technology needed to describe these two forms of gas clumping is quite
different.  In this section, we use only the halos that are well
resolved in the simulation as our sources ($m >2\times 10^9~
M_{\odot}$), and we set the luminosity proportional to the halo
mass. While this source prescription is probably unrealistic, we found
in \S \ref{feedback} that including less massive halos does not change
considerably the structure of reionization.

\subsection{IGM clumping}
 \label{clumping} We cannot realistically calculate the clumpiness of
the gas from the N-body simulation used in this paper.  In order to
investigate the effect of the clumping, we consider four toy models
for clumping of the IGM.  Simulation C1 uses a $256^3$ grid, setting
the baryonic overdensity to zero and the subgrid clumping factor
$C_{\rm cell}$ to unity in every cell.  In other words, the IGM is
completely homogeneous in this model. Simulation C2 is a $256^3$
simulation also with $C_{\rm cell} = 1$, but it uses the gridded
N-body density field.  The cell mass in C2 is $2 \times 10^9
~M_\odot$, approximately the Jeans mass for $10^4$ K gas at $z =
6$. Simulation C3 is a $512^3$ simulation with $C_{\rm cell} = 1$. The
cell mass in C3 is $3 \times 10^8 ~ M_\odot$, below the Jeans mass at
relevant redshifts, but possibly above the filtering mass.  Table 1
lists the specifications used in the C1--4 simulations.

When the Universe becomes reionized, the filtering mass $M_f$ can be
orders of magnitude smaller than the Jeans mass.  It takes hundreds
of millions of years for the gas to respond fully to the
photo-heating and clump at the limiting scale. Therefore, the
$512^3$ run is closer to reality than the $256^3$ one, but still
underestimates the effect of clumping on the IGM.  To account for
this higher degree of clumping, we run simulation C4.  This
is a $256^3$ simulation with twice the ionizing
efficiency of the other runs such that overlap occurs at around the
same time.  In addition, we set the subgrid clumping factor in C4 to
\begin{eqnarray}
C_{\rm cell} &=& 1 + \frac{\rho_0^2}{\rho_{\rm
cell}^2} \,\int_{0}^\infty \, \frac{k^2 dk}{2 \pi^2} \, \left[ 1 -
W^2_{{\rm cell}}(k) \right] \, P_{\delta \delta}(k, z) \nonumber \\
&\times& \exp(-k^2/(k_f^{-1} + k_5^{-1})^{-2}),\label{eqn:Ccell}
\end{eqnarray}
where $k_f$ is the scale that contains the mass $M_f$ (which is given
by equation \ref{eq:fm}), $W_{{\rm cell}}$ is the cell window
function, and $k_5$ is the wavevector that corresponds to $10^5
~M_{\odot}$ at the mean density -- the minimum mass baryonic clump that
we allow, consistent with a minimal amount of reheating. For
simplicity, we use a spherical top hat in real space that has the same
volume as a grid cell for $W_{{\rm cell}}$.  We use the Peacock and
Dodds power spectrum for $P_{\delta \delta}(k, z)$.  The
filtering mass $M_f$ depends on the redshift at which the cell was
ionized.  Once a region is ionized, this mass increases with time and
$C_{\rm cell}$ typically decreases.

 Equation (\ref{eqn:Ccell}) would be correct if the window function of
a cell were instead a top hat in $k$-space, if mode coupling were
absent between modes smaller and larger than the cell scale, and if
the quantity $M_f$ were appropriate outside of linear theory (there is
evidence that it is appropriate [\S \ref{gnedin}]). Since we are considering
non-linear scales, mode coupling \emph{is} important and tends to
make the more massive cells have higher clumping factors than equation
(\ref{eqn:Ccell}) predicts.  In the limit in which most of the density
fluctuations are at scales smaller than the cell, equation
(\ref{eqn:Ccell}) predicts that the number of recombinations ($
\propto C_{\rm cell} \, \rho_{\rm cell}\,
\rho_{\rm cell}$) is independent of the cell's density. This
prediction is probably unphysical.
 
Note that we assume that the gas clumping in a cell is independent of
the cell's ionization fraction in all of the simulations.  This
assumption is justified for the gas in the diffuse IGM because this
low density gas stays almost fully ionized when an ionization front
passes, provided that there is an ionizing background.  Virialized
objects, such as minihalos, in which the local ionized fraction can be
a function of density, are included in the computation in \S
\ref{minihalos}.

The reionization scenarios in this section reach $\bar{x}_i = 0.5$
near $z = 7$.  The reionization epoch in simulation C4 is slightly
more extended than the other scenarios owing to an enhanced number of
recombinations.  The volume-averaged clumping factor in ionized
regions $C_V$ is $30$ at $z= 7$ in C4, whereas it is $1.6$ in C2 and
it is $2.7$ in C3. The total number of IGM photons that escape into
the IGM per ionized baryon is $3$ in C4 at the end of reionization,
whereas it is between $1.2-1.3$ in C2 and C3.  (The recombination rate
is proportional to the clumping factor.)  Note that we have removed
the particles that are associated with halos from the density grid in
these simulations since most absorptions within these halos are
already encapsulated in the factor $f_{\rm esc}$.  Other studies left
the halos in the density field \citep{gnedin00, ciardi03-sim},
yielding a large number of recombinations within the source cells
(which can be at hundreds of times the mean density) and therefore a
larger photon to ionized baryon ratio.

\begin{figure}
{ \epsfig{file=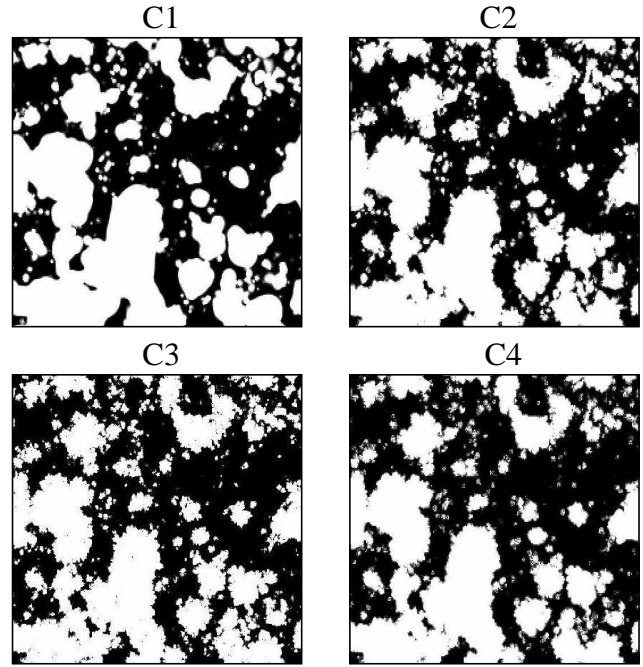, width=8.5cm}} \caption{The impact of gas clumping
on the structure of reionization.  A slice through the C1 ({\it top
left}), C2 ({\it top right}), C3 ({\it bottom left}) and C4 ({\it
bottom right}) runs at $z \approx 7$ and $\bar{x}_i \approx 0.5$.  All
the cells in C1 are at the the mean density. Simulation C2 is run on
top of the N-body simulation density field gridded to $256^3$, and C3
is the same but gridded to $512^3$.  Simulations C1, C2, and C3 set
$C_{\rm cell} = 1$.  Simulation C4 uses the $256^3$ grid with equation
(\ref{eqn:Ccell}) for $C_{\rm cell}$.  The additional clumpiness in
C2-C4 over C1 adds structure to the ionization front. Simulation C4
has at least $10\times$ more recombinations than in the other runs.}
\label{fig:clumpmaps}
\end{figure}

Figure \ref{fig:clumpmaps} depicts a slice through the box at
$\bar{x}_{i,V} \approx 0.5$ for the C1, C2, C3 and C4 simulations. The
ionization field in the top left panel (a snapshot from simulation C1,
which uses a homogeneous density field) has less structure on the
bubble edges than in the other runs.  The picture seen in the top left
panel is the most similar of all the panels to the picture of
reionization seen in Monte-Carlo realizations of HII regions using the
\citet{furlanetto04a} analytic model (see figures in \citet{zahn06}).
This model spherically averages the density field around a cell to
determine its ionization fraction, washing out much of the small-scale
structure in the density field.  The $256^3$ and $512^3$ runs have a
similar morphology despite the $512^3$ run's higher resolution and
larger volume-averaged clumping factor.  When we boost the subgrid
clumping factor substantially for the C4 run, this action does not
significantly change the morphology, even though this simulation has a
factor of $10$ more recombinations than the C2 and C3 simulations.

The $\Delta_{xx}^2$ for the C1--4 runs at $\bar{x}_{i,V}
\approx 0.5$ agree to better than $20\%$ at scales
larger than a Mpc.  Simulation C1 has the least power of all
the runs at megaparsec scales because it is missing the
density-induced structure at the bubble edges. In conclusion, the
differences in power from clumping in the considered models are minor
compared to the differences that arise owing to the different source
prescriptions.

In the C4 run, the subgrid clumping factor decreases as a function of
the cell's density (eqn. \ref{eqn:Ccell}).  We know that overdense
regions form substantially more structure, and, therefore, it is
possible that the subgrid clumping factor actually increases with
density.  To test whether such a prescription for clumpiness alters
the morphology of reionization, we ran a small-scale clumping run C5
with $C_{\rm cell} = 4 + 3 \,\delta_C$, in which $\delta_C$ is the
baryonic overdensity smoothed at the cell scale.  This clumping
prescription yields a similar scaling with density to the $C_{\rm
cell} \sim \rho_{\rm cell}^{0.86}$ that \citet{kohler05} finds in a
$4~ h^{-1} ~\Mpc$ simulation in which the halo particles are also
removed from the density field. This parametrization results in a
photon to ionized baryon ratio of $\approx 2$ at the end of
reionization and $C_{V} \approx 20$ throughout reionization. We do not
plot the results for C5, but we find that the HII regions have
slightly more structure on the edges in this case than in C3 and
C4. Overall, the structure of reionization is not significantly
altered in C5 from the other clumping runs.

Why does clumping not affect the large-scale morphology of
reionization?  Qualitatively, large-scale density fluctuations
significantly enhance the mass in sources that are present within an
overdense region relative to the mean.  However, the number of
absorptions and recombinations per unit volume are not enhanced by the
same margin.  This leads to the enhanced abundance of ionizing photons
winning in overdense regions and shaping the morphology of
reionization.  For a more quantitative treatment, one can solve for
the overdensity that a region must have to be ionized given some
source prescription and parametrization of the gas clumping.  This
overdensity threshold can then be used to calculate the bubble size
distribution with the excursion set formalism \citep{furlanetto04a,
bond91}.  For reasonable parameterizations of the clumping factor,
this exercise shows that clumping does not significantly change the
bubble morphology for fixed $\bar{x}_i$ \citep{mcquinn06b}.

On smaller scales, density fluctuations become more important in
shaping reionization.  For a single HII region ionizing a region of
$10 ~\Mpc$ in radius at $z=7$, the HII region is not a perfect sphere,
but has fluctuations in radius with $\Delta R/R \approx 0.2$.  These
fluctuations are generated by column density fluctuations between
different lines from the source to the bubble edges.  Lines with lower
column densities will lead to fingers protruding from the HII regions.
Such features are also present when many sources are within a bubble.

In addition to imprinting structure
on the bubble edges, clumpiness has a considerable effect on the part
in $10^4$ fluctuations in the neutral fraction within the bubbles.  We
will come back to this in future work.

In conclusion, quasi-linear density fluctuations imprint substructure
on the bubble edges, but do not affect the large-scale morphology of
the bubbles.  Quasi-linear fluctuations also increase the number of
recombinations and can extend reionization.  We address the effect of
self-shielding, non-linear density enhancements in \S \ref{minihalos}.

\subsection{Minihalos}
\label{minihalos}  The minimum mass minihalo that retains gas depends on
the thermal history of the IGM.  The Jeans mass at $z = 10$ for gas
that cools adiabatically since thermal decoupling from the CMB is $6
\times 10^3 ~M_\odot$ \citep{barkana02} and the filtering mass is
approximately ten times \emph{larger} \citep{gnedin98}. However,
reheating by X-rays prior to reionization will make the gas warmer
than this, erasing gas density fluctuations at progressively larger
scales. \citet{furlanetto06} estimates that if POPII stars are
responsible for reionization then the gas temperature is a couple
hundred degrees Kelvin prior to the time the Universe is $10\%$
ionized.  This estimate is based on extrapolating local X-ray
luminosities to high redshifts. A heated, neutral IGM has a Jeans mass
of $M_{\rm J} = 4 \times 10^6 ~ M_{\odot} \, [{T}/(200 \, {\rm K})
\times (1+z)/(10)]^{3/2}$.

An isolated minihalo that holds onto its gas during reheating will
subsequently lose its gas via photo-evaporation as ionizing flux
impinges upon it \citep{barkana99, shapiro03}. The timescale for
photo-evaporation $t_{\rm ev}$ of a minihalo is roughly the
sound-crossing time of the halo, which for $10^4$ K gas ionized is
\citep{shapiro03}
\begin{equation}
   t_{\rm ev} = 100 \,{\rm Myr} ~ \left( \frac{M}{10^7 \, M_{\odot}} \right)^{1/3}
   \,  \left ( \frac{10}{ 1 + z} \right).
\label{eq:evTime}
\end{equation}
This formula works well when the incident flux is large, but
under-predicts the evaporation time for the ionizing fluxes that are
typical during reionization \citep{iliev-mh}.  The duration of
reionization in our simulations is a few hundred million years,
comparable to the evaporation timescale of minihalos (eqn.
\ref{eq:evTime}), suggesting that minihalos will be present for
all times during reionization.

Prior to evaporation, a minihalo is optically thick for a typical
ionizing photon.  An incident photon ionizes a hydrogen atom
within the minihalo and the photon's energy is converted primarily
into kinetic energy of the minihalo gas rather than into additional IGM
ionizing photons. The mean free path at $z= 6$ to intersect a halo
of mass $(10^5, 10^6, 10^7)$ $M_\odot$ within a virial radius is
$(4, 7 ,17) ~\Mpc$ [or at $z= 12$ is $(6, 19 ,74)~ \Mpc$]
if we assume the Press-Schechter mass function.

Several previous calculations have attempted to encapsulate the effect
of minihalos via a clumping factor (e.g., \citealt{haiman00}).  We
emphasize that this is not an appropriate way to treat
minihalos. Minihalos are self-shielded such that the densest inner
regions should not contribute to the clumping \citep{iliev-mh}. In
addition, in the context of large-scale simulations, only a small
portion of photons through a cell will intersect a
minihalo. \citet{ciardi06} was the only previous study to investigate
minihalos in the context of large-scale radiative transfer
simulations.  However, \citet{ciardi06} set the cell optical depth in
minihalos to be the average optical depth for all sight-lines through
the cell. The average optical depth from minihalos can be large even
though the vast majority of sight-lines will not intersect a
minihalo. A more appropriate model for the minihalos is to treat them
as dense absorbers with an absorbing cross section $\sigma_{\rm mh}$.
We adopt this treatment for the minihalos: Only the fraction
$\sigma_{\rm mh}/L_{\rm cell}^2$ of photons in a ray that passes
through a cell of side-length $L_{\rm cell}$ are absorbed in a
minihalo of cross section $\sigma_{\rm mh}$ that sits within the cell.

We add minihalos to our simulation box using the mean value method,
Method 1 discussed in \S \ref{mergertree}.  We use the Press-Schechter
mass function for the minihalos, but using the Sheth-Tormen mass
function instead would not affect our conclusions.  The mass function
of minihalos is calculated in each cell on a $64^3$ coarse grid, and
the mass in mininhalos for a coarse cell is divided equally among its
fine cells. This method is justified because the mean free path for
photons is always larger than the width of a coarse cell in our
models.

In all of our calculations, we assume that once a region is ionized,
no new minihalos form owing to ``Jeans mass suppression''. To
incorporate this suppression, we calculate the opacity of a cell at
redshift $z$ that was ionized at $z_{\rm rei}$ from the mass
$n_{PS}(m, \delta_{0,M}, M_c, z_{\rm rei})$ rather than $n_{PS}(m,
\delta_{0,M}, M_c, z)$.  However, we find that our results are
unchanged if we omit suppression. This is because minihalos are
abundant at the redshifts relevant to our study such that the number
density of minihalos is not rapidly changing. For higher redshift
reionization scenarios, the degree to which minihalos are suppressed
from forming in ionized regions can play a larger role
\citep{ciardi06}.

To understand the impact of minihalos, we adopt three simplified
models for these objects. In our most extreme model for minihalos
(simulation M3), we make all minihalos with mass greater than $10^5 ~
M_{\odot}$ opaque to ionizing photons out to a virial radius.  The
mass cutoff of $10^5 ~M_{\odot}$ is consistent with a minimal amount
of reheating.  Simulation M2 is the same as M3, except that the
effective cross section $\sigma_{\rm mh}$ of a minihalo to ionizing
photons is not fixed as a function of time, but instead the function
used for $\sigma_{\rm mh}$ is motivated by the evolution of the cross
section in the simulations presented in \citet{shapiro03} -- initially
the outer layers of the gas in minihalos are quickly expelled leaving
a dense core, which is evaporated over a time $t_{\rm ev}$. The
formulas we use in M2 for $\sigma_{\rm mh}$ and $t_{\rm ev}$ are
presented in \S \ref{iliev-fits} along with a discussion of potential
drawbacks. Finally, simulation M1 has the same sources as the other
minihalo runs but does not include any
minihalos.\footnote{\citet{barkana02} finds that minihalos impose a
much shorter mean free path than in our models.  The reason for this
difference is because \citet{barkana02} uses a static model for the
minihalos, which results in each minihalo having a much larger cross
section.  \citet{shapiro03} finds that the outskirts of the minihalo
are quickly photo-evaporated, leaving a smaller cross section than in
\citet{barkana02}. The parameterizations in this section assume the
outskirts are quickly evaporated.}

\begin{figure}
\rotatebox{-90}{\epsfig{file=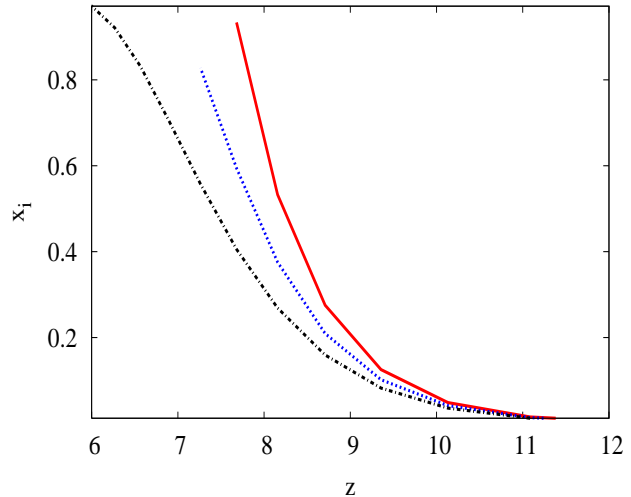, width=7cm, height=8.7cm}}
 \caption{The volume-averaged ionization fraction for simulations M1
({\it solid curve}), M2 ({\it dotted curve}), and M3 ({\it dash-dotted
curve}).  In M3 the minihalos absorb more photons than in M2, and
there are no minihalos in M1.  All simulations have the same source
prescription.  The presence of minihalos extends the duration of
reionization.}\label{fig:ximini}
\end{figure}

Figure \ref{fig:ximini} plots the ionization history of simulations M1
({\it solid curve}), M2 ({\it dotted curve}) and M3 ({\it dash-dotted
curve}). All of these simulations use the source luminosity of
$\dot{N}(m) = 9\times 10^{49} \; m/(10^8 M_{\odot})$ photons s$^{-1}$. The
absorptions in the minihalos extend reionization by less than $100$
million years in M2 and by more than $250$ million years in simulation
M3. In addition, one in every two ionizing photons in M2 is destroyed
in a minihalo by $\bar{x}_{i,V} = 0.8$, and two in every three are
destroyed in M3.

\begin{figure}
\begin{center}
{ \epsfig{file=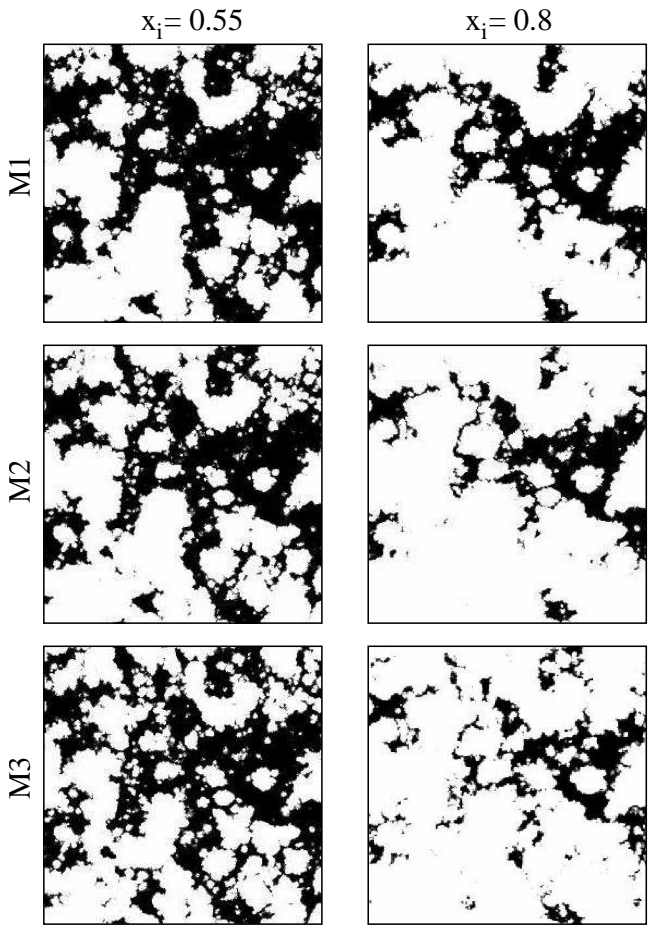, width=8.5cm}}
\end{center}
 \caption{The effect of minihalos on the ionization maps for
$\bar{x}_{i, V} \approx 0.55$ ({\it left panels}) and $\bar{x}_{i, V}
\approx 0.8$ ({\it right panels}).  The top panels are slices from the
M1 simulation in which minihalos do not affect the propagation of the
ionization fronts (because we have a limited number of outputs, the
top panels are at approximately a $7\%$ lower ionization fraction than
the others).  White regions are ionized and black are neutral.  The
middle panels are from M2, in which minihalos are evaporated with a
prescription motivated by the results of \citet{shapiro03} and
\citet{iliev-mh}.  The bottom panels are from simulation M3 in which
minihalos are not evaporated, and all halos above $10^5 ~M_{\odot}$
absorb ionizing photons with impact parameter less than one virial
radius (yielding an average photon mean free path of $4 ~\Mpc$).
Minihalos inhibit the largest bubbles from
growing.}\label{fig:minihalos}
\end{figure}

Figure \ref{fig:minihalos} shows slices through the M1, M2, and M3
simulations ({\it top, middle, and bottom panels, respectively}) at
$\bar{x}_{i,V} = 0.55$ ({\it left panels}) and at $\bar{x}_{i,V} =
0.8$ ({\it right panels}). [Note that, due to a limited number of
outputs at which to compare, the output for simulation M1 is $\approx
7\%$ less ionized than the outputs for the other runs.]  The total
number of absorptions inside minihalos increases from simulation M1 to
M2 to M3. The major effect from minihalo absorptions is that the
largest bubbles (bubbles larger than the photon mean free path) grow
more slowly, whereas the growth of the smaller bubbles is uninhibited.
This effect is particularly noticeable in simulation M3, in which the
average mean free path is $4 ~ \Mpc$.  The mean free path becomes
larger than this as the smallest halos are evaporated in simulation
M2, such that the effect of minihalos on the bubble sizes is less
significant. The smaller bubbles are still larger in M2 than in
M1. (Since M1 is at a $\approx 7\%$ smaller $\bar{x}_i$, if we
compared at the same $\bar{x}_i$, this trend would be more
noticeable.)

\begin{figure}
\rotatebox{-90}{\epsfig{file=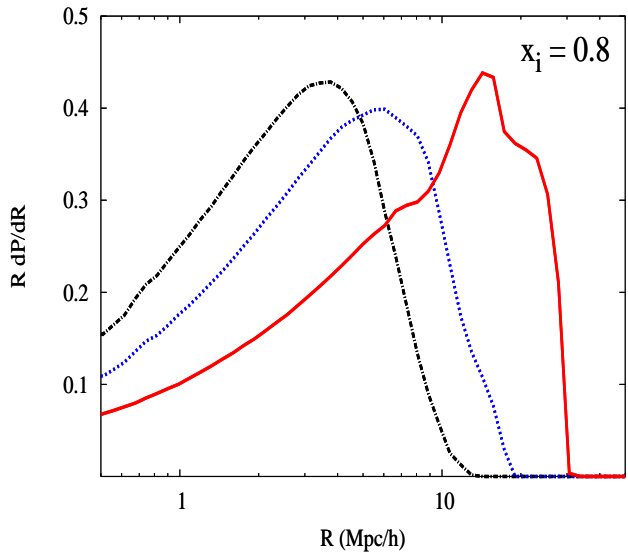, width=7.5cm, height=8.8cm}}
\caption{The bubble size distribution for the M1 ({\it solid curve}),
M2 ({\it dotted curve}), and M3 ({\it dot-dashed curve}) simulations
for $\bar{x}_i \approx 0.8$. For the minihalo models, the bubble size
peaks at roughly the photon mean free path.}
\label{fig:mini_bub}
\end{figure}

Figure \ref{fig:mini_bub} shows the bubble PDF for the minihalo runs,
in which the bubble radius is defined as in \S \ref{sources}. We
confirm that the bubbles are smaller when the minihalos are present,
particularly once the biggest bubbles become larger than the photon
mean free path.  At $\bar{x}_{i, V} = 0.8$, the characteristic bubble
radius is $20 ~\Mpc$ in M1 ({\it solid curve} in
Fig. \ref{fig:mini_bub}), $7~\Mpc$ in M2 ({\it dotted curve}) , and $4
~\Mpc$ in M3 ({\it dot-dashed curve}).  In the minihalo models, the
characteristic scale is set roughly by the average photon mean free
path, which is $4 ~\Mpc$ in simulation M1.  This decrease of the
characteristic bubble scale from the dense absorbers was first
predicted in analytic models \citep{furlanetto05}.  However, we do not
find the sharp cutoff in effective bubble size at the scale of the
mean free path found in the analytic work of \citet{furlanetto05}. The
reasons for this difference are primarily that analytic models make
the simplifying assumptions that the mean free path is spatially
uniform and that photons from a source cannot travel a distance
further than one mean free path.

\begin{figure}
\rotatebox{-90}{\epsfig{file=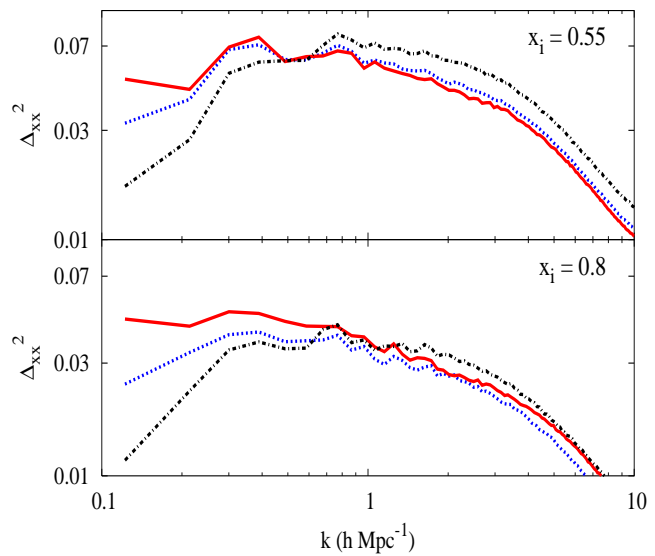, width=7.5cm, height=8.8cm}}
\caption{The ionization fraction power spectrum $\Delta_{xx}^2$ for
the M1 ({\it solid curves}), M2 ({\it dotted curves}), and M3 ({\it
dot-dashed curves}) simulations for $\bar{x}_i \approx 0.55$ ({\it top
panel}) and $\bar{x}_i \approx 0.8$ ({\it bottom panel}). (Note that
the snapshot from M1 is really at a $7\%$ lower ionization fraction
than the snapshots from M2 and M3 in both panels.) At fixed
$\bar{x}_i$, the minihalos suppress the large-scale ionization
fluctuations and increase the size of the fluctuations at smaller
scales.  }
\label{fig:pk_mini}
\end{figure}

Figure \ref{fig:pk_mini} plots $\Delta_{xx}^2$ for the M1 ({\it solid
curves}), M2 ({\it dotted curves}), and M3 ({\it dot-dashed curves})
simulations for $\bar{x}_{i,V} \approx 0.55$ ({\it top panel}) and
$\bar{x}_{i,V} \approx 0.8$ ({\it bottom panel}).  The minihalos
suppress the large-scale ionization fluctuations and increase the size
of the fluctuations at smaller scales.  The significance of the effect
of minihalo absorptions increases with ionization fraction as the
bubbles become larger.  Notice that the total power is contained
within the box for the models with minihalos in Figure
\ref{fig:pk_mini} (the power peaks at smaller scales than the box
scale) -- the presence of minihalos reduce the size of the box
necessary to simulate reionization.  Note that the differences in
$\Delta_{xx}^2$ among the minihalo models we consider (simulations
M1--3) are not as large as the differences in $\Delta_{xx}^2$ among
the source models for $\bar{x}_i = 0.55$ (simulations S1--S4, Fig.
\ref{fig:pksources}).  However, for larger ionization fractions ({\it
see bottom panel}) the effect of minihalos on the structure of
reionization can be comparable to the effect of different source
models.

Dense systems other than minihalos may limit the mean free path of
ionizing photons during reionization. \citet{gnedin00} finds that such
systems play an important role in reionization in
radiative-hydrodynamics simulations. The effect of these ``Lyman-limit''
systems should be similar to the effect we find for the minihalos.

\section{Redshift Dependence}
\label{redshift} Up to this point, we have only considered reionization
scenarios in which overlap occurs at $z \approx 7$ and result in $\tau
= 0.06-0.08$. However, WMAP's measurement of $\tau = 0.09 \pm 0.03$
does not rule out overlap at higher redshifts. Further, the popular
conclusion that quasar absorption spectra require that reionization is
ending at $z \approx 6.5$ is being hotly debated \citep{fan06,
mesinger04, wyithe04, lidz06, becker06, lidz06b}.  At higher
redshifts, there are fewer galaxies above $m_{\rm cool}$, enhancing
Poisson fluctuations, and the galaxies that do exist are more biased
on average. In addition, at higher redshifts the Universe is more
dense, resulting in a higher level of recombinations. Finally, at
higher redshifts the number of galaxies is growing more quickly,
possibly leading to a shorter duration for the reionization epoch.
Owing to all these differences, it is interesting to investigate how
the structure of reionization when comparing at fixed $\bar{x}_i$
changes with redshift. Analytic models predict that the bubble size
distribution at fixed $\bar{x}_i$ is relatively unchanged with
redshift \citep{furlanetto04a}

\begin{figure}
{
 \epsfig{file=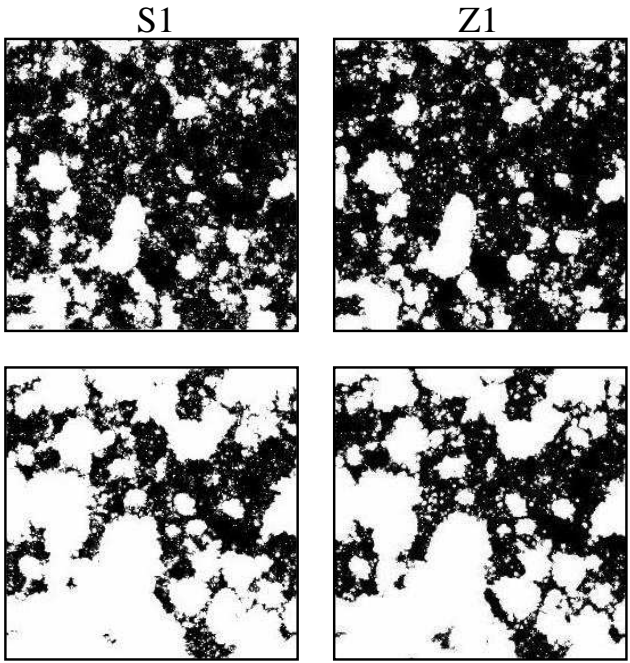, width=8.5cm}}
 \caption{The dependence of the morphology on the redshift of
reionization.  Slices through snapshots from the S1 and Z1
simulations.  Simulation Z1 has the same source prescription as S1, but
with five times the ionizing efficiency.  The top panels compare S1 at
$z = 8.2$ ({\it left}) and Z1 at $z = 11.1$ ({\it right}), both with
$\bar{x}_{i, V} \approx 0.3$, and the bottom panels compare S1 at $z =
7.3$ ({\it left}) and Z1 at $z = 10.1$ ({\it right}), both with
$\bar{x}_{i, V} \approx 0.6$.  The ionization fields from S1 and Z1
are strikingly similar.  }\label{fig:zdep1}
\end{figure}

Figure \ref{fig:zdep1} compares snapshots from the S1 simulation and
Z1 simulation, which has the same sources as S1, but where each source has
five times the ionizing efficiency.  The higher efficiency results in
reionization occurring earlier by a redshift interval of $\Delta z \approx 3$.
 The top panels compare S1 at $z = 8.2$ ({\it left}) with Z1 at $z = 11.1$
({\it right}), both with $\bar{x}_{i, V} \approx 0.3$.  The bottom panels
compare S1 at $z = 7.3$ ({\it left}) with Z1 at $z = 10.1$ ({\it right}), both
with $\bar{x}_{i, V} \approx 0.6$.  The ionization field is very
similar between S1 and Z1 for fixed $\bar{x}_i$.

\begin{figure}
{
 \epsfig{file=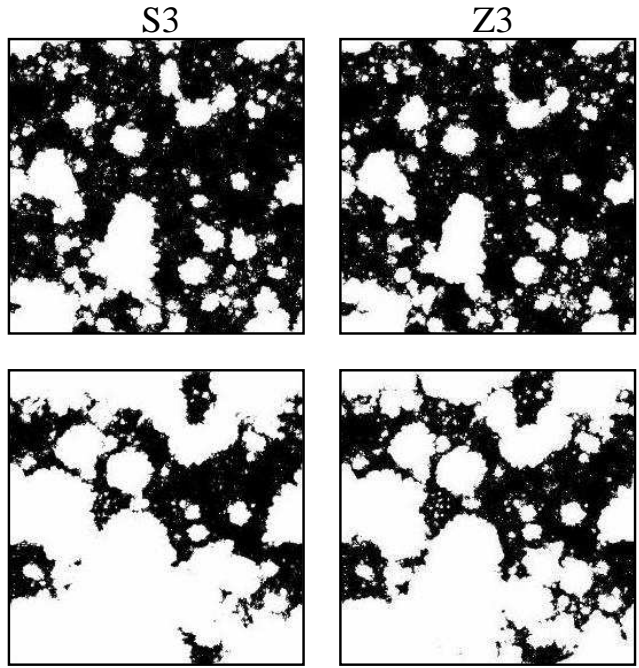, width=8.5cm}}
 \caption{Slices through snapshots from the S3 and Z3 simulations.
 Simulation Z3 has the same source prescription as S3, but with five times
the ionizing efficiency.  The top panels compare S3 at $z = 8.2$
({\it left}) and Z3 at $z = 11.1$ ({\it right}), both with $\bar{x}_{i, V}
\approx 0.3$, and the bottom panels compare S1 at $z = 7.3$ ({\it left})
and Z1 at $z = 10.1$ ({\it right}), both with $\bar{x}_{i, V} \approx 0.6$.  The
ionization field is very similar between S3 and
Z3.}\label{fig:zdep2}
\end{figure}

We also ran simulation Z3, which uses the same source prescription as
S3 ($\alpha \propto m^{2/3}$), except that the sources in Z3 are five
times as efficient as in S3. More massive sources dominate the
ionizing efficiency in the S3 and Z3 models than in S1 and Z1.  Since
the more massive sources are closer to the exponential tail of the
Press-Schechter mass function, the part of the mass function which is
rapidly changing, we might expect a more significant difference in the
ionization maps as we change the redshift of overlap than we found in
the previous case. Figure \ref{fig:zdep2} compares the ionization maps
for the S3 and Z3 simulations ({\it left and right panels,
respectively}).  The ionization maps are, as with S1 and Z1, very
similar.  The differences between the $\Delta_{xx}^2$ calculated from
S1 and Z1 (or from S3 and Z3) are $\lesssim 10\%$ at fixed
$\bar{x}_i$.

We can understand why the maps look so similar at fixed $\bar{x}_i$ by
again comparing the power spectra of the sources at these redshifts.
The top panel in Figure \ref{fig:mtps} shows the luminosity-weighted
power spectrum $\Delta_{hh}^2$ of the sources used the S1/Z1
simulations ({\it solid curves}) and S3/Z3 simulations ({\it dotted
curves}) at $z = 6.6$ ({\it thick curves}) and $11.1$ ({\it thin
curves}).  The differences between $\Delta_{hh}^2$ for the S1 (or S3)
sources at $z = 6.6$ and at $ z = 11.1$ are much smaller than the
differences between the $\Delta_{hh}^2$ for the S1, S3 and S4 source
models.  Therefore, we would expect the differences between the ionization
fields at fixed $\bar{x}_i$ but separated by $\Delta
z \approx 3$ to be smaller than the differences between the fields for
the S1, S3, and S4 models, which is what we find.

Because the ionization maps do not depend strongly on the redshift of
reionization, we expect that our conclusions in previous sections hold
for slightly higher redshift reionization scenarios.  The invariance of the
ionization fields with redshift also implies that the conclusions in
this paper are not sensitive to the value of $\sigma_8$.  If
reionization occurs at very high redshifts, redshifts where the
cooling mass sources are extremely rare, then the topology of
reionization will shift from the topology seen in S1 to something
closer to what is seen in S4 -- the bubbles will become larger and
more spherical (see discussion in \citet{zahn06}).

\section{Observational Implications}
\label{observations} In this section, we briefly discuss the
potential of observations to distinguish different reionization
models.  We limit the discussion to Ly$\alpha$ emitter surveys and
21cm emission.  In future work, we will discuss the implications for
these and other observations in more detail.

\subsection{Ly$\alpha$ Emitter Surveys}
\label{lyalpha}

Narrow band Ly$\alpha$ emitter surveys are currently probing redshifts as high
as $z = 6.5$, and projects are in the works to search for higher
redshifts Ly$\alpha$ emitters \citep{kashikawa06, santos04b, barton04, iye06}.
If there are pockets of neutral gas at these redshifts, the statistics
of these emitters can be dramatically altered \citep{madau00,
haiman02, santos04, furl-galaxies05, malhotra06}.  Sources must be in
large HII regions for the Ly$\alpha$ photons to be able to redshift
far enough out of the line center to escape absorption. Therefore,
the structure of the HII regions will modulate the observed properties of
the emitters.  Because of this modulation, Ly$\alpha$ emitters could be
a sensitive probe of the HII bubbles during reionization. From the
current datat on these emitters, there is disagreement as to whether
there is evidence for reionization at $z = 6.5$ \citep{kashikawa06,
malhotra06}.

The calculations in this section are all at $z=6.5$, the highest
redshift at which there are more than a handful of confirmed
Ly$\alpha$ emitters.\footnote{The redshifts that can be probed from
the ground are limited by sky lines, which contaminate a significant
portion of the relevant spectrum. At $z = 6.5$ there is a gap in
these lines that allows for observations.} Rather than re-run our
simulations to generate maps with different ionization fractions at $z
= 6.5$, we instead use the property that the structure of HII regions
at fixed $\bar{x}_i$ is relatively independent of the redshift (as
demonstrated in \S \ref{redshift}). We take the ionization field from
the simulation for higher $z$ and use this field in combination with
the $z = 6.5$ sources. Since the photo-ionization state of the gas
within an HII bubble is dependent on the redshift, we remove the
residual neutral fraction within each HII region when calculating the
optical depth to absorption $\tau_{{\rm Ly}\alpha}$.  The residual
neutral gas primarily affects the blue side of the line, which we
assume is fully absorbed.  We also ignore the peculiar velocity field
in this analysis.  The peculiar velocities are typically much smaller
than the relative velocities due to Hubble expansion between the
emitter and its HII front, and, therefore, this omission does not
affect our results.

Next, we integrate the opacity along a ray perpendicular to the front
of the box from each source to calculate $\tau_{{\rm
Ly}\alpha}$. Rather than assume an intrinsic Ly$\alpha$ line profile
and follow many frequencies, we calculate the optical depth
$\tau_{{\rm Ly}\alpha}$ for a photon that starts off in the frame of
the emitter at the line center $\nu_0$ and set the observed luminosity
$L_{\rm obs, Ly\alpha} = a \, L_{\rm int, Ly\alpha} \,
\exp[-\tau_{{\rm Ly}\alpha}(\nu_0)]$, in which $a$ is a constant of
proportionality that encodes the amount of absorption at the line
center. For reference, an isolated bubble of $1$ proper Mpc that is
fully ionized in the interior has $\tau_{{\rm Ly}\alpha}(\nu_0) =
1$. We also assume the escape fraction is independent of halo mass
such that $L_{\rm int, Ly\alpha} = b \, \dot{N}_{\rm
UV}$.\footnote{The precise value of the proportionality constants $a$
and $b$ does not matter for the subsequent discussion in this section.
The value of $a$ and $b$ does matter if we are to compare our results
with observations.  The standard assumption is that $a = 0.5$ (the
blue side of the line is absorbed while the red side is unaffected),
but $a$ is probably smaller than this value \citep{santos04}. In
principle, we could calculate $a$ from the ionization field in our
simulation, but we leave this to future work. In the absence of dust,
$b = 0.67 \, (1 - f_{\rm esc}) h \nu_{\alpha}$ \citep{osterbrock89}
such that if we assume $f_{\rm esc} \ll 1$ then the observed
Ly$\alpha$ luminosity of these sources is $L_{\alpha} = 3 \times
10^{41} m/(10^{10} \, M_{\odot})$ erg s$^{-1}$ in simulation S1 for $a
= 0.1$. The observed emitters have luminosities of $2 \times 10^{42} -
1 \times 10^{43}$ erg s$^{-1}$, which correspond to halos with $m
\gtrsim 10^{11} ~ M_{\odot}$ in S1.  Presently, surveys cover $\sim
10^{6} ~ \Mpc^3$ at $z = 6.5$, but probe only the $\sim 100$ brightest
emitters in that volume \citep{kashikawa06}.  Assuming for simplicity
that all halos host an emitter (which is certainly not true in
detail), we reproduce the observed abundance of Ly$\alpha$ emitters
$\bar{n} \approx 2 \times 10^{-4}~\Mpc^{-3}$ \citep{kashikawa06} if
all halos with masses $\gtrsim 3 \times 10^{11} \, M_{\odot}$ host
observed emitters (assuming $\sigma_8 = 0.9$).}  In future work, we
will do a more thorough analysis that includes the velocity field, the
neutral fraction within the bubbles, as well as several frequencies
around $\nu_0$.  We also ignore here any stochasticity in the
Ly$\alpha$ emission from galaxies.  \citet{santos04} discusses the
importance of many of the effects that are ignored in the calculations
in this section.

\begin{figure}

 \rotatebox{-90}{\epsfig{file=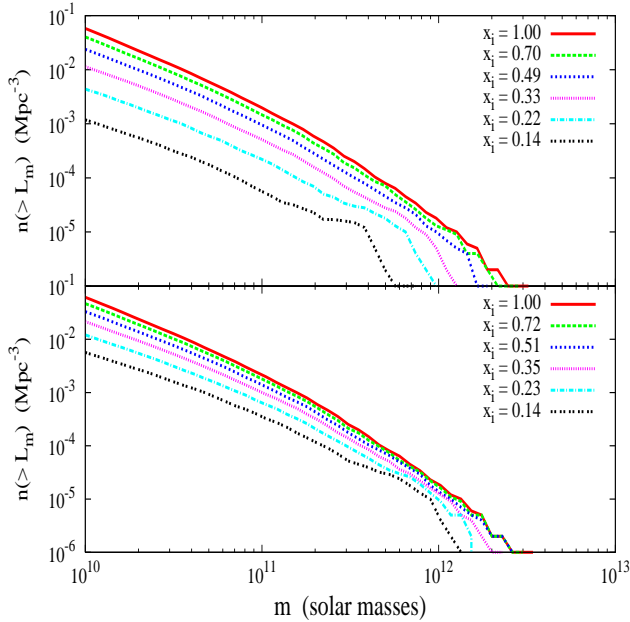, width=8cm, height=8.7cm}}
\caption{The number density of Ly$\alpha$ emitters above a certain
Ly$\alpha$ luminosity $L_m = a L_{\rm int, Ly\alpha}$ at $z = 6.5$ and
for several different volume averaged ionization fractions (denoted by
$x_i$ in the plot). Here we use the fact that there is monotonic
relationship between luminosity and mass in our source models, which
allows us to plot halo mass on the abscissa. The top panel is from the
S1 simulation in which $L_{\rm int, Ly\alpha} \propto m$ and the
bottom is from S3 in which $L_{\rm int, Ly\alpha} \propto
m^{5/3}$. Because the sources in S3 are rarer, the bubbles are larger
and, therefore, the luminosity function is less suppressed.  Current
surveys at $z = 6.5$ probe a volume of $10^6~ \Mpc^3$ and have found
$\approx 100$ emitters.}
\label{fig:lumfuncs}
\end{figure}

Figure \ref{fig:lumfuncs} plots the number density of Ly$\alpha$
emitters with luminosity above $a L_{\rm int, Ly\alpha}$ for several
volume-averaged ionization fractions denoted by $x_i$ in the plot. We
use the fact that there is monotonic relationship between luminosity
and mass in our models, which allows us to plot mass on the
abscissa. The top panel is from S1 in which $L_{\rm int, Ly\alpha}
\propto m$ and the bottom is from S3 in which $L_{\rm int, Ly\alpha}
\propto m^{5/3}$.  Because the ionizing sources in S3 are rarer, the
bubbles are larger and the luminosity function is less suppressed. For
both simulations, once the Universe is more than half ionized, the
luminosity function is not significantly suppressed at fixed
$\bar{x}_i$.  The normalization of the luminosity function is very
sensitive to ionization fractions $\bar{x}_i \lesssim 0.5$ in both
models.

The luminosity function for different ionized fractions in our
calculations is suppressed from the intrinsic luminosity function by a
factor that is fairly independent of halo mass
({Fig. \ref{fig:lumfuncs}}). This prediction for the observed
luminosity function is similar to the analytic predictions of
\citet{furl-galaxies05}, which use a similar source prescription to
that of S1.  However, the luminosity function we predict is less
suppressed by a factor of $1.5-2$. This small difference is partly
because \citet{furl-galaxies05} underestimates the free path a photon
will take inside a bubble.  In \citet{furl-galaxies05}, for
computational convenience the distance for a photon to travel within a
bubble is defined as the distance from the source to the
\emph{nearest} neutral clump rather than the distance along the ray to
the bubble edge.

\citet{kashikawa06} finds significant evolution in the
luminosity function between $z=5.7$ and $z = 6.5$ and suggests that
this might be evidence for reionization. However, the $z = 6.5$
luminosity function differs most with the $z =5.7$ at the high mass
end, as opposed to our prediction of it being uniformly suppressed. We
suggest that the observed evolution is more consistent with
cosmological evolution in the abundance of massive host halos, rather
than reflecting an evolving ionized fraction.

\begin{figure}
\begin{center}{\epsfig{file=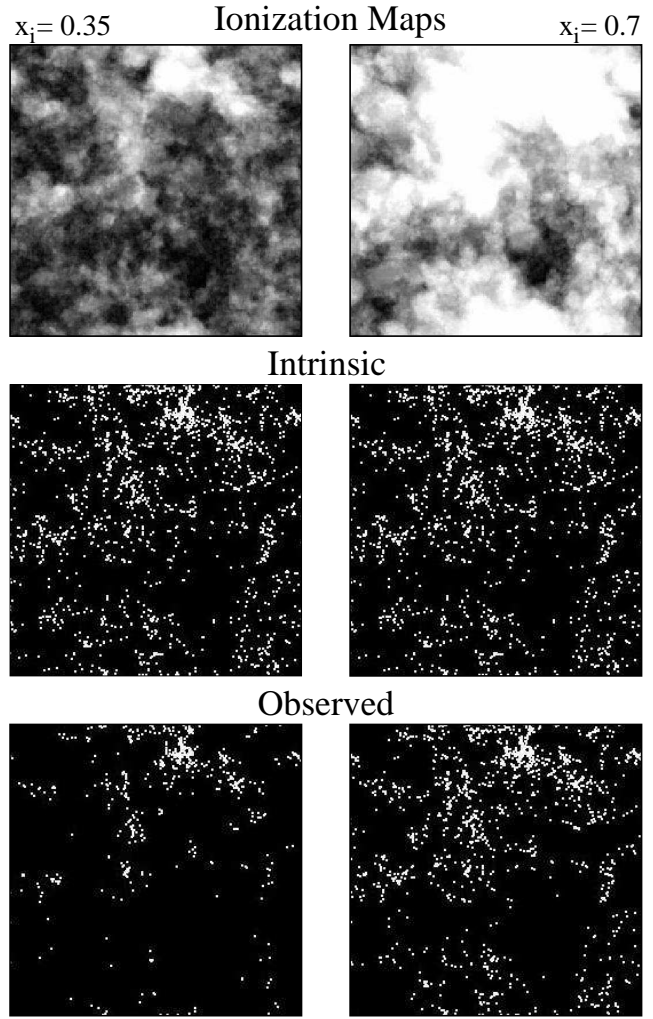, width=8.5cm}}
\end{center}
\caption{Mock $z = 6.5$ Ly$\alpha$ emitter surveys at two different stages of
reionization.  The left panels are for $\bar{x}_{i,V} = 0.35$ and the
right are for $\bar{x}_{i,V} = 0.7$. The panels would
subtend $0.6$ degrees on the sky. All panels use the S1 simulation.
{\it Top Panels}: A map of the average ionization fraction from a slice
through the box of width $31 ~\Mpc$. In white regions the projection
is fully ionized and in black it is fully neutral. {\it Middle
Panels}: The intrinsic population of Ly$\alpha$ emitters in the same
$31 ~\Mpc$ slice.  Our mock survey is sensitive to emitters with halo
masses $m > 5 \times 10^{10} ~M_{\odot}$. There are $1800$ such halos
in the survey, resulting in a density that is an order of magnitude
higher than the density probed by current Ly$\alpha$
surveys. {\it Bottom Panels}: The observed distribution of Ly$\alpha$
emitters if the Universe is ionized as given in the corresponding top
panels.  The observed distribution of these emitters is modulated by
the presence of HII regions (see top panels for location of HII
regions and contrast with intrinsic distribution).  }
\label{fig:lymaps}
\end{figure}

Figure \ref{fig:lymaps} shows maps of the Ly$\alpha$ emitters in
simulation S2 with $m > 5\times 10^{10} ~ M_{\odot}$.
This mock survey would subtend $0.6$ degrees on the sky and has a
volume of $3\times10^5~ \Mpc^3$.  The left panels are for
$\bar{x}_{i,V} = 0.35$ and the right are for $\bar{x}_{i,V} =
0.7$. The top panels show the average ionization fraction for a
projection of width $31 ~\Mpc$, corresponding to a narrow band filter
with width $\Delta \lambda = 100$ angstroms. White regions are fully
ionized and black are fully neutral. The middle panels show the
intrinsic population of Ly$\alpha$ emitters. There are $1800$ of these
halos in the survey; the density of these halos is an order of
magnitude higher than the density currently probed by narrow band
Ly$\alpha$ surveys. The bottom panels show the observed emitters [with
observed luminosity greater than $L_{{\rm int}, Ly\alpha}(m = 5\times
10^{10} ~M_{\odot})$], which is modulated by the ionization field in
the top panel. In the left, bottom panel, there are 500 visible
emitters and in the right, bottom there are 1400.  Detecting these
large-scale variations in the abundance of Ly$\alpha$ emitters would
be a unique signature of patchy reionization.  In future work, we
calculate several clustering statistics from our emitter maps.

Current surveys at $z=6.5$ are dominated by Poisson fluctuations and
are not yet sensitive to density fluctuations or bubble-induced
fluctuations from reionization.  Figure \ref{fig:lymaps} illustrates,
however, that once surveys resolve enough sources then they will be
able to detect fluctuations from the HII regions (these fluctuations
are generally much larger than the density-induced fluctuations).  At
larger scales, the bubble fluctuations start to dominate over the
Poissonian fluctuations. The fluctuations generated by the bubbles are
order unity at the bubble scale, whereas the intrinsic Poisson
fluctuations are given by $\Delta(k)^2 = k^3/ (2\pi^2 \, \bar{n})$.
If we equate this with unity at $k = 0.5 \,h \;\Mpc^{-1}$
(corresponding to a bubble scale of $R = 18 \, \Mpc$), we find that
surveys must be sensitive to source densities of $\bar{n} =
2\times10^{-3} \; \Mpc^{-3}$ in order to overcome Poisson noise and be
able to image the bubble-induced fluctuations.  The requirements for a
statistical detection are less stringent.  Currently, surveys can
probe to densities of $\bar{n} \approx 2\times 10^{-4} \;\Mpc^{-3}$.
In future work, we provide a more quantitative estimate for the number
density that surveys must probe to detect reionization.

In future work we will also include the effect of minihalos and gas
clumpiness on the Ly$\alpha$ emitters.  Minihalos/Lyman-limit systems
limit the bubble size and so could potentially suppress the observed
luminosity function more substantially than we find in the S1 and S3
simulations.

\subsection{21cm Emission}
\label{21cm} The LOFAR and MWA radio interferometers are being built
to observe high redshift neutral hydrogen via the 21cm line, and the
GMRT interferometer can already observe at these wavelengths. These
telescopes hope to observe an increase in brightness temperature
over that of the CMB at wavelengths $\lambda = 21{\rm cm} \, (1+z)$
for $z > z_{\rm rei}$ with amplitude
\begin{eqnarray}
T_{21}(\nhat, z) &=& 26 \, \left(1 - x_i(\nhat, z) \right) \,
\left(1 + \delta_b(\nhat, z)\right) \nonumber\\
& \times & \left(\frac{T_s(\nhat, z) - T_{\rm
CMB}(z)}{T_s(\nhat, z)}
\right)\, \left(\frac{\Omega_B \, h^2}{0.022} \right) \nonumber \\
& \times & \left( \frac{0.15}{\Omega_m \,h^2}\, \frac{1+z}{10}
\right)^{1/2} ~{\rm mK}, \label{eqn:T21}
\end{eqnarray}
where $T_s$ is the spin temperature and $\delta_b$ is the baryonic
overdensity.  Equation (\ref{eqn:T21}) (as well as our calculations)
neglects redshift-space distortions, which can enhance the signal
\citep{barkana04a}.  However, these distortions offer only a small
enhancement of the signal on the large scales of interest at which
ionization fluctuations dominate the signal
\citep{mcquinn06}.  We also assume $T_s \gg T_{\rm CMB}$ in this
section, likely a good approximation during the bulk of the
reionization epoch \citep{ciardi03-21cm, furlanetto06}.

\begin{figure}
\rotatebox{-90}{ \epsfig{file=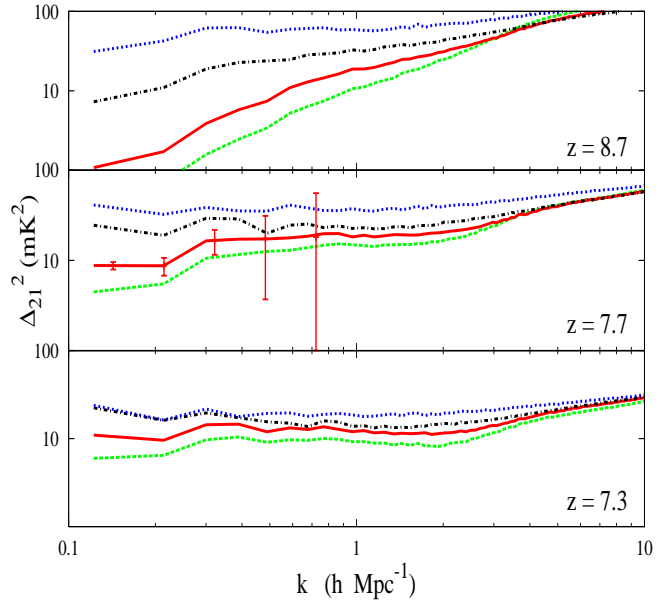, width=8cm, height=8.8cm}}
\caption{The 21cm power spectrum for the S1 ({\it solid curves}), S2
({\it dashed curves}), S3 ({\it dot-dashed curves}), and S4 ({\it
dotted curves}) simulations [$\Delta_{21}^2(k) = k^3 \langle
T_{21}(k)^2\rangle /2\pi^2$].  For the top panels, $\bar{x}_{i, V}
\approx 0.2$ ($\bar{x}_{i, M} \approx 0.3$), for the middle panels,
$\bar{x}_{i, V} \approx 0.5$ ($x_{i, M} \approx 0.6$), and for the
bottom panels, $\bar{x}_{i, V} \approx 0.7$ ($\bar{x}_{i, M} \approx
0.8$).  In S2 the lowest mass sources dominate (with masses $m \sim m_{\rm
cool}$), and in S4 the highest mass sources dominate ($m \sim 5 \times
10^{10} ~M_{\odot}$). At scales $k \lesssim 1 \, h~ \Mpc^{-1}$,
$\Delta_{21}^2$ scales approximately as $\Delta_{xx}^2$, such that the
21cm signal is a sensitive probe of the bubble structure. The error
bars are the detector noise plus cosmic variance errors on the power
spectrum for MWA, assuming $1000$ hours of integration and a bandwidth
of $6 ~ \MHz$. Foregrounds will eliminate the sensitivity to the
signal for $k \lesssim 0.1 ~ h\; \Mpc^{-1}$.}
\label{fig:21cmpk}
\end{figure}

Figure \ref{fig:21cmpk} plots the 21cm power spectrum for the S1 ({\it
solid lines}), S2 ({\it dashed lines}), S3 ({\it dot-dashed lines}),
and S4 ({\it dotted lines}) simulations for $\bar{x}_{i,V} = 0.2$
({\it top panel}), $\bar{x}_{i,V} = 0.5$ ({\it middle panel}), and
$\bar{x}_{i,V} = 0.7$ ({\it bottom panel}). The S3 and S4 simulations
have much more power at large scales than the other runs, particularly
at early times owing to the larger bubbles in these runs
(Fig. \ref{fig:sourcesbub}). The signal is very flat on the scales
probed by the box for most $\bar{x}_i$.  If we had a larger box, a
sharp decline in power would be observed at larger scales than the
bubbles.

The power spectra in Figure \ref{fig:21cmpk} do not include
absorptions from minihalos.  If minihalos are as abundant in reality
as they are in models M2 and M3, this will suppress significantly the
large-scale power in the 21cm signal (see the $\Delta_{xx}$ in
Fig. \ref{fig:pk_mini}).  The effect of the minihalos on
the 21cm power spectrum is qualitatively different from the effect of
changing the sources and should also be observable.

The projected 1-$\sigma$ errors for MWA for a $1000$ hour observation
in a 6 MHz band in bins of width $\Delta k = 0.5 \,k$ are shown in the
middle panel in Figure \ref{fig:21cmpk}.  The sensitivity of LOFAR is
comparable to that of MWA. The details of this sensitivity calculation
are discussed in \citet{mcquinn06}.  Because of foregrounds,
experiments will encounter difficultly detecting smaller $k$-modes
than are plotted here \citep{mcquinn06}.  The first generation of
interferometers are most sensitive to $k$ greater than $0.1 \,h \;
\Mpc^{-1}$ and less than $1 \, h \;\Mpc^{-1}$.  MWA and LOFAR should
be able to distinguish between the S1, S3 and S4 reionization
scenarios at a fixed ionized fraction.  If we marginalize over the
ionized fraction, it is unclear whether MWA can still distinguish
between these models.  The second generation 21cm experiments MWA5000
and SKA will be at least $10\times$ more sensitive than MWA and LOFAR
\citep{mcquinn06}.

In addition, it might be possible to use the {\it evolution} of the
21cm signal to separate models.  For the models considered in this
paper, the duration of reionization is fairly short, spanning an
interval of $\delta z = 2-4$.  It is quite possible that upcoming 21cm
experiments will be able to observe the entire breadth of this epoch.
The duration of reionization is shortest if the largest mass sources
dominate the ionizing budget.  Also, minihalos tend to cause a delay
at the end of reionization ({ Fig. \ref{fig:ximini}}).  Perhaps
combining information on the duration of reionization with the power
spectrum at different times can help us understand the source
properties as well as the role of the minihalos.  Higher order terms
in the 21cm power spectrum may aid in distinguishing different
reionization scenarios \citep{lidz06c}. \citet{zahninprep} (in
preparation) investigates how well upcoming 21cm experiments can
constrain certain reionization models.

\section{Conclusions}

We have run a suite of $94$ Mpc radiative transfer simulations to
understand the size distribution and morphology of HII regions for
$0.1 < \bar{x}_i < 0.8$. These simulations are the first that include
sources down to the cooling mass and that are large enough to contain
many HII regions.  We have incorporated structures that
all large-scale simulations of reionization do not resolve with
analytic prescriptions.

We find that the morphology of HII regions is most sensitive to the
parameter $\bar{x}_i$.  If we compare different reionization scenarios
at the same $\bar{x}_i$, they tend to look similar.  This is not to
say other factors besides $\bar{x}_i$ do not change the morphology.
The sources responsible for reionization are the second most important
factor.  If we compare at fixed $\bar{x}_i$, we find that the HII
regions become larger (by as much as a factor of $4$) and more
spherical as the sources become rarer.  The bubbles are larger for the
rarest sources because these sources are the most biased.

The next most important factor for shaping the morphology is the
presence of minihalos.  Once the mean free path for a photon to
intersect a minihalo becomes smaller than the bubble size, the effect
of minihalo absorptions becomes important.  As a result, minihalos inhibit the
largest bubbles from growing.  If we use the results of
\citet{shapiro03} and \citet{iliev-mh} to characterize the minihalos,
we find that these objects have a modest effect on the overall
properties of the HII regions at fixed $\bar{x}_i$, decreasing
$\Delta_{xx}^2$ by as much as $50\%$ for the largest modes in our box.
In a more extreme case we considered, in which the average mean free path is
$4~\Mpc$ during reionization, the impact of minihalos is even more
substantial.  Minihalos do not have the same effect as changing the
source efficiency.

We find that thermal feedback and quasi-linear density inhomogeneities
have more minor consequences for the topology of the bubbles at fixed
$\bar{x}_i$.  This is fortunate because these quantities are poorly
constrained.  Feedback does not substantially change the morphology of
reionization at fixed $\bar{x}_i$ because the bias difference between
the $m >10^8 ~M_{\odot}$ halos and the $m >10^9 ~M_{\odot}$ halos is
relatively small.  (The typical halo that is suppressed by feedback is
located in a similar region as the typical halo which is not.)
Megaparsec-scale, quasi-linear density fluctuations add structure to
the boundaries of the HII regions.  This additional structure is
ignored in analytic models.  However, as we increase the level of
small-scale gas clumping, either by increasing the resolution or by
increasing the subgrid clumping factor, the large-scale structure of
the HII regions is largely unaffected at fixed $\bar{x}_i$.  This is
true even if small-scale gas clumping results in a substantial number
of recombinations.  We find that the $\Delta_{xx}^2$ at fixed
$\bar{x}_i$ differ by no more than $20\%$ as we vary the volume
averaged clumping factor from $0$ to $30$.  The qualitative reason why
clumping does not affect the morphology of reionization at fixed
$\bar{x}_i$ is because the enhanced photon production in a large-scale
overdense region (that is a bubble) is always able to overcome the
enhanced number of recombinations, even in extreme clumping models.

The conclusions in this paper hold if overlap occurs at slightly
higher redshifts then in our typical simulation in which $z_{\rm
overlap} \approx 7$.  In fact, we found that if we boosted the source
efficiencies such that at $z_{\rm overlap} \approx 10$, the ionization
map is essentially unaffected.  We showed that this can be explained
by the relatively small differences in the luminosity-weighted source
power spectrum at $z =7$ compared to that at $z=10$ in the models we
considered.  The conclusion that the structure of reionization does
not depend on the redshift is no longer true if we compare with
simulations that reionize at much higher redshifts, redshifts at which
the sources become extremely rare. In this case, reionization may
have a similar morphology to simulation S4, in which the rarest
sources dominate.

In this paper, we did not concentrate on predicting the duration of
reionization.  However, many of the effects we consider impact the
duration of reionization, even if they do not impact the morphology.
We find that our most extreme minihalo model extends the duration of
reionization by $250$ million years ($\Delta z \approx 1.5$).  In
addition, feedback on POPII-like ionizing sources from photo-heating can in
extreme cases extend reionization by $200$ million years.

Analytic models provide a convenient and intuitive framework to
understand the structure of reionization \citep{furlanetto04a,
furl-models, zahn06}. These models do not suffer from the same scale
limitations as simulations, and they supply a quick method to explore
the large parameter space relevant to reionization.  In addition,
these models enhance our physical intuition regarding the processes
that shape this epoch.  We have confirmed the analytic predictions
that the bubble size distribution is approximately log-normal and that
the sizes of the bubbles increase as the sources become more
biased. Further, we confirm the prediction of analytic models that
bubble sizes are largely unchanged if we compare the same model at
different redshifts, yet fixed ionized fraction.  We also showed,
however, that current analytic models encounter some difficulties in
describing the effect of minihalos and of Poisson fluctuations in the
source abundance on the structure of reionization.  Analytic models
cannot incorporate the sophisticated models for thermal feedback, gas
clumping, and minihalo evaporation that it is possible to include in
radiative transfer simulations.

Upcoming observations have potential to distinguish the source models
we considered.  We make predictions for the luminosity function of
Ly$\alpha$ emitters as a function of $\bar{x}_i$.  We construct maps
of Ly$\alpha$ emitters from a mock survey that show large-scale
fluctuations in the distribution of emitters due to the HII regions,
suggesting that future measurements of the clustering of emitters may
reveal the signature of patchy reionization.  Future 21cm arrays hold
much promise for probing reionization; measurements of the power
spectrum with the MWA and the LOFAR arrays can distinguish the S1, S3
and S4 source models.

Upcoming observations can reduce the parameter space that reionization
simulations need to explore. If we can measure the number of ionizing
photons produced by high mass galaxies and bright quasars at high
redshifts, this will reduce the almost total freedom we currently have
in the ionizing luminosity.  Observations of the mean free path of
ionizing photons at high redshifts may reveal whether the Lyman-limit
systems are the minihalos as well as how fast these systems are being
evaporated.  A precise measurement of the Thomson scattering optical
depth from the CMB will constrain the average redshift of
reionization.

It is important to continue to improve large-scale simulations of
reionization to understand the reionization process in more detail.
Future simulations need to investigate the effect of more realistic
star formation models, metal pollution, and alternative sources of
ionizing photons. In addition, larger simulations than are presented
here are necessary to statistically describe this epoch for $\bar{x}_i
\gtrsim 0.7$.  It is also useful to run small-scale simulations to
develop more realistic subgrid parameterizations for the minihalos and
for the clumping factor.  These parameterizations will be essential
for modeling the end of reionization, a time when the rate of
evaporation of the Lyman-limit systems plays a key role in determining
the structure of reionization.  Also, such parameterizations are
necessary to extend our calculations to model accurately the part
in $10^4$ neutral fraction fluctuations that characterize the high redshift
Ly$\alpha$ forest.

An accurate interpretation of future observations of reionization,
while certainly challenging, does not appear impossible.  The
characteristics of HII regions during reionization might have depended
on a huge number of poorly constrained parameters, making it
impossible to interpret observations of this epoch.  We find that this
is not the case.  The morphology of the HII regions at fixed
$\bar{x}_i$ boils down primarily to the properties of the sources and
of the minihalos/Lyman-limit systems.

\section{acknowledgments} We thank Mark Dijkstra for helpful conversations
regarding thermal feedback and Ly$\alpha$ emitters, Volker Springel
for providing his Lean Gadget--2 code, Roman Scoccimarro for providing
his 2LPT code, and Steven Furlanetto for useful comments on the
manuscript.  MM would like to thank the NSF graduate student
fellowship.  The authors are also supported by the David and Lucile Packard
Foundation, the Alfred P. Sloan Foundation, and NASA grants
AST-0506556 and NNG05GJ40G.

\begin{appendix}
\section{Radiative Transfer Algorithm}
\label{code} For the simulations in this paper, we employ the
\citet{sokasian01, sokasian03, sokasian04} cosmological radiative
transfer code, but with several significant changes that are discussed
below.  This algorithm inputs grids of the density field as well as a
list of sources and then casts rays from every source, randomizing the
order of the sources within this loop.  Radiative feedback on the
density field is ignored. Rays are split adaptively using the HealPIX
algorithm \citep{abel02} such that, at a minimum, $N$ rays from a
source intersect every cell face (for this paper, we set $N =
2.1$). This algorithm does not iterate the ray casting within each
time step to converge to the correct ionized fraction in each
cell. Instead, once a cell has been ionized by a source within a time
step, rays from other sources will pass through it. This
simplification allows for the algorithm to process more sources and
larger volumes than other codes.  In the limit of few sources and few
time steps, this simplification can lead to artificial structure in the
HII regions.  However, with the vast number of sources in the
simulations in this paper, even with relatively coarse time steps we
choose, this artificial structure is minimized (as we will demonstrate
later in this section).

The temperature history of the gas is not tracked by this code.
Instead, the code assumes that ionized regions are at $T = 10^4$
Kelvin. The temperature affects the number of recombinations in the
simulation because $\alpha_B \propto T^{-0.7}$, as well as the
detailed photo-ionization state of the gas within the HII regions.  The
analysis we have done in this paper does not depend on the
photo-ionization state of the gas. In addition, the value for the
subgrid clumping factor, which determines the number of recombinations
with in a cell, is highly uncertain, such that we would not gain any
precision from including a full calculation of the IGM
temperature.

What follows is a list of the important modifications that we have
made to the original \citet{sokasian01} algorithm:

\begin{itemize}
\item  Previously, cells were either ionized or neutral.  Cells can now be fractionally ionized.  We assume that the
ionizing front is paper thin such that each cell can be broken up
into a neutral part and an ionized part.
\item Each ray holds a number of photons.  In the original
\citet{sokasian01} algorithm, the first ray that hit a cell from a
particular source carried all the information that the cell needed
about the source. Subsequent rays from the same source did not affect
the cell. The advantage to having each ray contain a specific number
of photons is that it is trivial to conserve photons as well as to
include photon sinks.  The disadvantage is that the ionization front
has a numerical width that is wider than in the previous algorithm. We
find that the width of the front in the new algorithm is approximately
$2$ cells for a single source.  The thickness of the front is smaller
than $2$ cells in the limit relevant to this paper of many faint
sources.
\item The orientation of the HealPIX ray casting scheme is randomly
rotated between each time step, and the order with which the rays are
initially cast is also randomized. When rays split adaptively, the
order is again randomized over the daughter rays.  All of this
randomization is done to minimize artifacts owing to the order in which
operations are performed.
\item
Once a ray has traveled a distance equal to $\eta \, R_{\rm
box,proper} \, [3 \dot{N}_{source}/(4 \pi \dot{N}_{\rm box})]^{1/3}$,
it can no longer split into daughter rays, where $\dot{N}_{\rm
source}$ is the ionizing luminosity from the source and $\dot{N}_{\rm
box}$ is the total luminosity of all the sources in the box.  We set
$\eta = 5$ for this paper.  Until this distance, rays split adaptively
such that a set number of rays intersect every cell face. This
simplification is justified by the large numbers of sources in an
HII region (typically more than $1000$ sources), making it unnecessary
to have rays from one side of an HII region cover the entire front on
the other side.  Our approximation results in the correct fluxes in
the cells in the limit of many sources.  We have investigated
quantitatively whether this simplification makes a difference in the
ionization maps.  The middle panel in Figure \ref{fig:convergence}
plots the cross correlation coefficient at two times between a
simulation with no ray termination and a simulation with the
prescription for ray termination used in this paper (see the caption in
Fig.  \ref{fig:convergence} for the definition of the cross
correlation coefficient).  There is essentially no difference between
the maps. This simplification results in the algorithm running over a
factor of $5$ faster at high ionized fractions.
\item
The previous algorithm reset the density in each cell after a time
step to the density field in the next snapshot, but did not change the
ionized fraction in the cell to account for the dynamics of the
ionization field.  For example, a cell that becomes fully ionized
would remain fully ionized in subsequent time steps (neglecting
recombinations), even if it gained neutral material from a neighboring
cell during these time steps. This resulted in the total number of
ionized atoms not being conserved by the previous algorithm between
time steps.  To remedy this issue in the current algorithm, we account
for a dynamic density field by assigning some ionization fraction to
each particle in the N-body simulation and then re-gridding the
ionization map between time steps to account for the particle
dynamics. We suspect that other cosmological radiative transfer
algorithms performed on top of a static density grid ignore the
dynamics of the ionization field in their computations.\footnote{Note
that our code still ignores thermal feedback and therefore does not
capture the full dynamics of the gas.}
\item
N-body particles that are associated with halos
are not included in the density field used by the radiative transfer
algorithm.  Otherwise, cells with sources would have substantial
overdensities, and ionizing photons from within the cell would have to
ionize these cells prior to escaping into the IGM.  These absorptions
are already counted in the escape fraction.  Removing the halo
particles from the gridded density field is also appropriate for rays
incident on this cell. The gas in the massive source halos has cooled
to form a small disk that is much smaller than the cross section of
the cell. Therefore, the vast majority of photons coming from exterior
to the cell do not intersect this disk.  The gas within galaxies
during reionization absorbs a negligible amount of external photons
because the mean free path of these photons to intersect a galaxy is
large (larger than the $94$ Mpc box size employed in this paper).
\end{itemize}

\begin{figure}
\begin{center}
\rotatebox{-90}{
 \epsfig{file=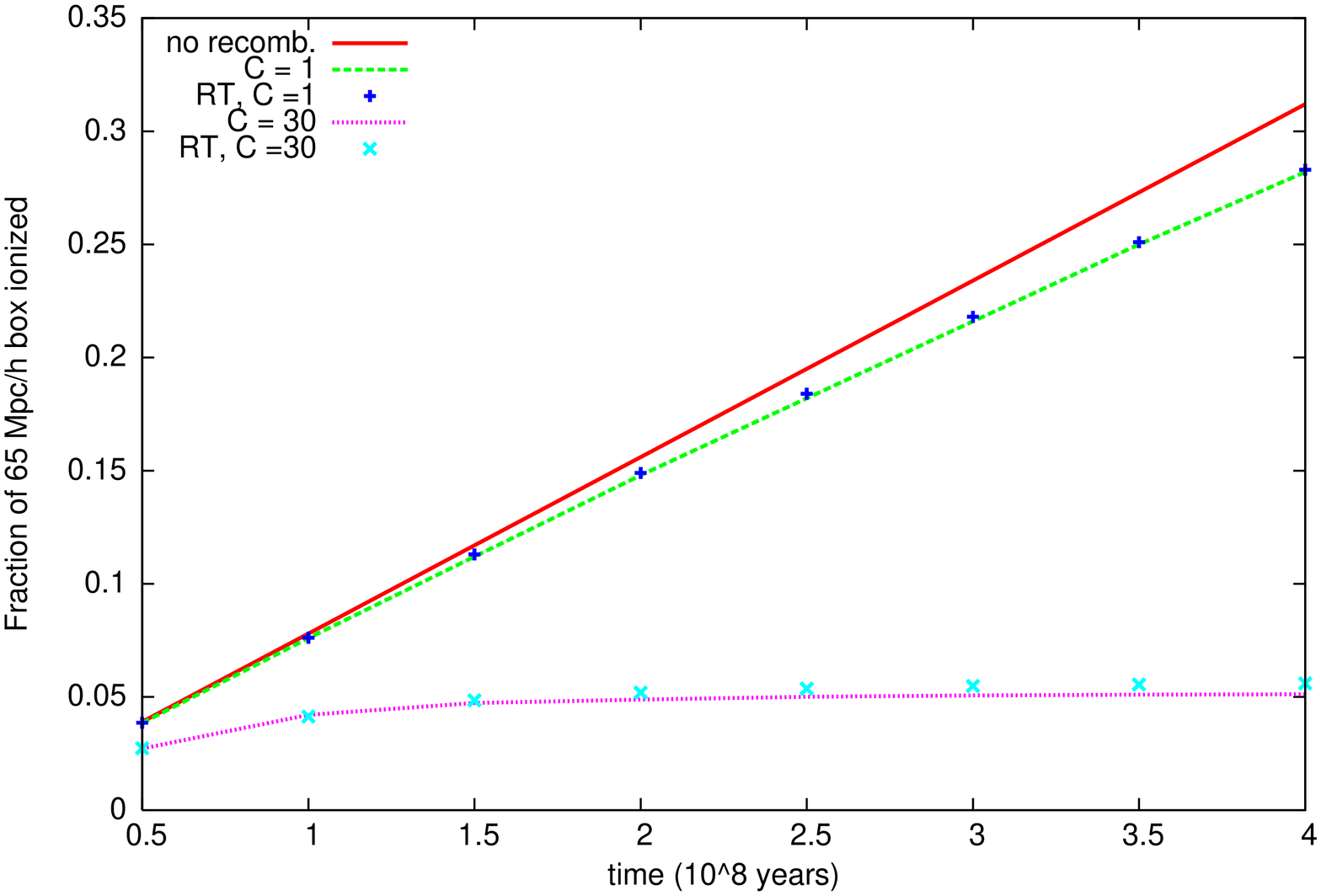, height = 8.6cm, width=7.3cm}}
 \end{center}
 \caption{
The fraction of the simulation box that is ionized by a single
source with $\dot{N} = 10^{56} ~s^{-1}$.  This calculation is
performed at $z = 6$ with all cells at the mean density of the
Universe and with the subgrid clumping factor $C$ equal to $1$ and $30$.
The solid, dashed and dotted curves are from a theoretical
calculation for no recombinations, recombinations with $C = 1$ and
recombinations with $C= 30$, respectively.  The pluses and crosses
are the radiative transfer algorithm with $C = 1$ and $C = 30$ with
coarse time steps of $\Delta t = 5 \times10^7$ years.  The radiative
transfer agrees well with the theoretical result, slightly
under-predicting the number of
recombinations.}\label{fig:recomb_test}
\end{figure}

\begin{figure}
\begin{center}
\rotatebox{-90}{ \epsfig{file=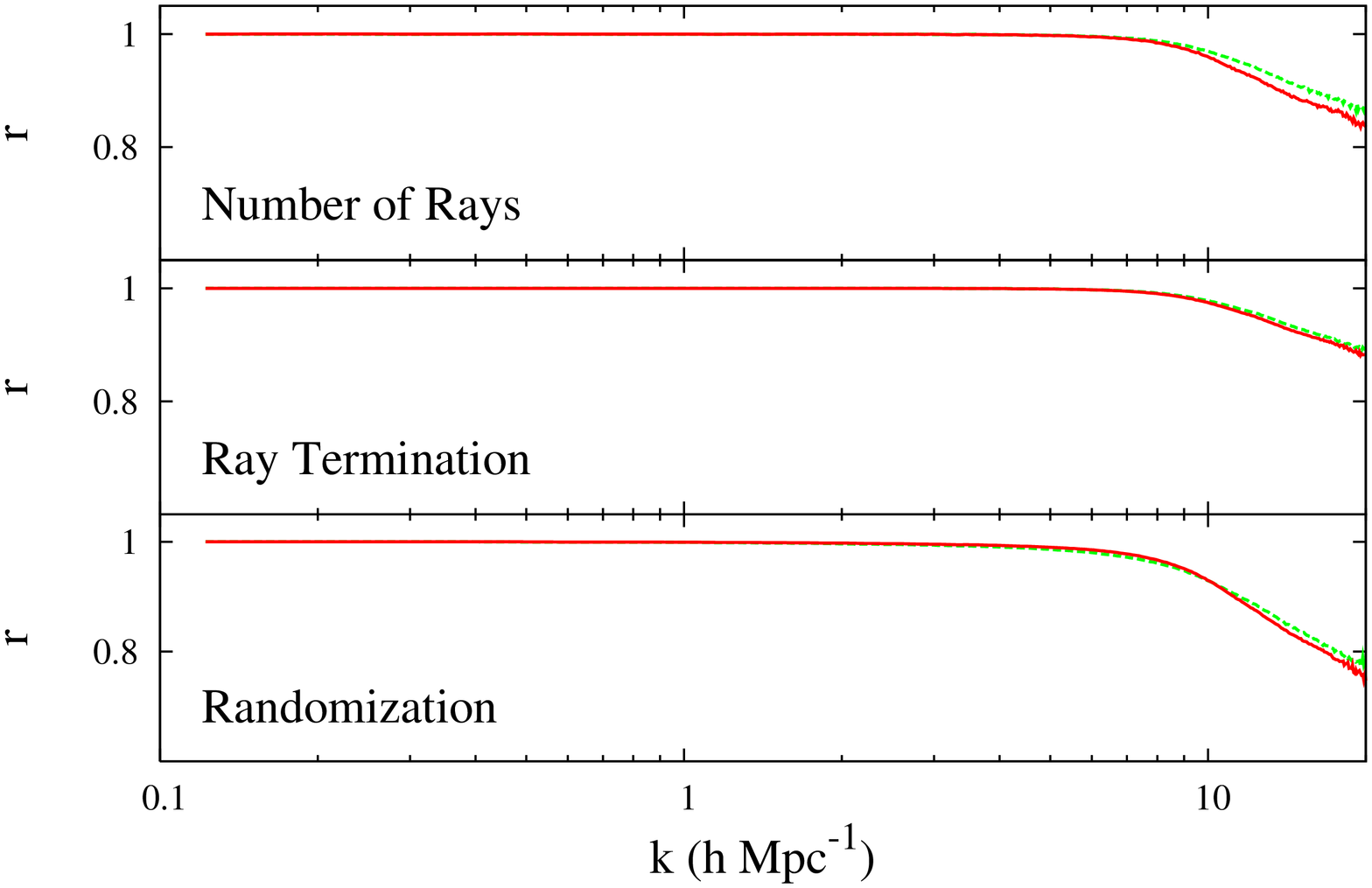, height=8.8cm, width=8cm}}
\end{center}
\caption{The cross correlation coefficient $r = P_{x_1 x_2}(k)/
\sqrt{P_{x_1 x_1} (k) \, P_{x_2 x_2} (k)}$ between the ionization maps
from different simulations.  All panels are run with $\dot{N}(m) =
3\times10^{49} \; m/(10^8 ~M_{\odot})~ s^{-1}$ and using only halos with
$m > 2 \times 10^9 ~ M_{\odot}$.  The cell scale is at $k \approx 20
~h \; \Mpc^{-1}$.  {\it Top Panels}: The cross correlation coefficient
for a simulation that casts $64$ times the number of initial rays and
ensures that $2$ times more rays intersect every cell as in the
fiducial runs. The solid curve is for $\bar{x}_i = 0.2$ and the dashed
is for $\bar{x}_i = 0.5$.  {\it Middle Panels}: The cross correlation
coefficient of a simulation that terminates rays with the condition
given in this section versus a simulation with no ray termination. The
solid curve is for $\bar{x}_i = 0.3$ and the dashed is for $\bar{x}_i
= 0.5$.  The maps are extremely similar even though the simulation without ray
casting took five times longer to reach $\bar{x}_i = 0.5$.  {\it
Bottom Panels}: Cross correlation coefficient for two runs with two
different sets of random numbers. The random numbers set the order of
the sources for ray casting. The solid curve is for $\bar{x}_i = 0.2$
and the dashed is for $\bar{x}_i = 0.7$. All panels
show that the ionization field has converged sufficiently in our simulations.}
\label{fig:convergence}
\end{figure}

We subjected the radiative transfer algorithm to several tests.  As a
simple test, we put one source with $\dot{N} = 10^{56} \, {\rm
photons~} {\rm s}^{-1}$ in a $65.6$ Mpc/h box with $256^3$ cells, with
each cell at the mean density of the $z =6$ Universe, and set the
clumping factor $C$ in each cell to $C = 1$ or $C = 30$. In Figure
\ref{fig:recomb_test}, we compare the fraction of the box that is
ionized in this test to the fraction that is predicted by theory
(using coarse time steps of $5\times 10^7$ years). Even with such
coarse time steps, this algorithm matches the theory curves well.

Because the algorithm does not iterate to find the ionized fraction,
this might lead to artificial structure if the time step is too
coarse. In the limit of an infinitely small time step, this algorithm
gives us the exact solution.  In this paper we use a time step of $50$
million years. To test convergence, we run two cosmological
simulations on the $256^3$ grid (which we label simulation 1 and 2),
using the halos with $m > 2\times 10^9 ~M_{\odot}$ as our
sources. Each simulation uses a different set of random numbers to
establish the order of the sources for ray casting. If the 50 million
year time step is too coarse, the ionization maps from these two
simulations would differ substantially, whereas the time step is
sufficiently small if the ionization maps differ insignificantly.  The
bottom panel of Figure \ref{fig:convergence} plots the cross
correlation coefficient $r = P_{x_1 x_2}/ \sqrt{P_{x_1 x_1} \, P_{x_2
x_2}}$ between these two runs for ionization fractions of $\bar{x}_i =
0.2$ ({\it solid curves}) and $\bar{x}_i = 0.7$ ({\it dashed
curves}). The cross correlation coefficient is close to unity at most
scales, dropping to $0.8$ at the cell scale.  Note that the cross
correlation coefficient is a stringent test.  We have also looked
at the power spectrum of these runs. The power spectrum of the ionized
fraction differs negligibly between these two runs, differing by about
$0.3\%$ at $k = 10 ~ h ~ \Mpc^{-1} $ and $1.5\%$ at $k = 20 ~h
~\Mpc^{-1}$.

We have also investigated whether the ionization structure seen in our
fiducial runs has converged to the true structure by either increasing
the number of rays that are cast or by increasing the mesh size to
$512^3$. First, we ran with a simulation that casts a much larger
number of rays per source than the fiducial number of rays ($64\times$
more initially and with the criteria that a minimum of $4.1$ rather
than $2.1$ rays intersect every cell).  We find that the fiducial
number of rays is enough to capture the structure of the HII region
({\it top panel}, Fig. \ref{fig:convergence}).  Second, we have run a
resolution test, comparing a higher resolution $512^3$ radiative
transfer simulation without recombinations to a $256^3$ run without
recombinations.  We re-grid the $512^3$ snapshots to $256^3$ resolution
for comparison.  We find that for all $\bar{x}_i$ the power spectra
have converged to within $10\%$ at scales with $k < 10 ~h \;
\Mpc^{-1}$.  The agreement is even better then this for $\bar{x}_i >
0.5$.

The computation speed of this code scales in a time step as $N_r \,
R^2 \, N_s$ where $N_r$ is the number of rays through each cell, $R$
is the characteristic bubble size, $N_s$ is the number of sources.
When the bubbles are large, the code slows down considerably. We are
working on ways to ameliorate this issue with the code.  Most
simulations in this paper took less than two days to reach
$\bar{x}_i = 0.8$ on a single $2.2$ GHz AMD Opteron processor.

\section{Fitting Formula for Minihalo Evaporation}
\label{iliev-fits} \citet{iliev-mh} provide fitting formula for the
evaporation of minihalos by POPII stars. These simulations do full
radiative hydrodynamics on minihalos, which are modeled prior to front
crossing as truncated isothermal spheres (TIS) with self similar
infall.  They provide the formula for the evaporation timescale
\begin{eqnarray}
t_{\rm ev} &=& 150 \, \left (\frac{M}{10^7 ~ M_\odot} \right)^{0.434}
\, F^{-0.35 + 0.05 \, \log_{10}(F)} \nonumber \\ 
& &\times \left[0.1 + 0.9 \,
\left(\frac{1 +z}{10} \right) \right]  ~ {\rm Myr},
\label{eqn:tevIliev}
\end{eqnarray}
where $F$ is the flux (which is time-independent in their
simulations).  To apply these formula to simulation M2, we use for $F$
the time averaged flux incident on a cell, with averaging starting
after the cell becomes ionized.

In Figure 29 in \citet{shapiro03}, the effective cross section of a
halo for absorbing an ionizing photon as a function of time is
plotted for a $10^7~ M_{\odot}$ halo. \citet{iliev-mh} does not
provide parameterized fits to the effective cross section, which we
need in our calculations.  To proceed, we fit by eye the curve for
the effective cross section in \citet{shapiro03}.  We find the
function
\begin{equation}
\frac{\sigma_{\rm mh}}{\pi \, r_t^2} = \frac{1}{3} \times 10^{-1.7
\, \left(\frac{t}{t_{\rm ev}}\right)^{1.5}}, \label{eqn:shapcross}
\end{equation}
where $r_t = 0.754 \, [M/(10^7 \, M_{\odot})]^{1/3} \, 10/(1 + z)$ is
the scale radius for the TIS profile.  By construction in the
simulations of \citet{shapiro03}, $\sigma_{\rm mh} =\pi \, r_t^2$ at
$t = 0$.  However, on a timescale of order a million years the
outskirts of the halo are evaporated, consuming a meager amount of
photons.  The denser inner regions take a significantly longer time to
evaporate.  We set $\sigma_{\rm mh} =\pi \, r_t^2$ for the first $5$
million years, and subsequently use equation (\ref{eqn:shapcross}) in
run M2.  Of course, the function we use for $\sigma_{\rm mh}$ likely
does not scale correctly with redshift or halo mass. We anticipate
that it over-predicts the cross section for halos with $m <10^7~
M_{\odot}$, since the gas in the outskirts of
these smaller halos will be easier to evaporate.

\section{Filtering Mass}
\label{gnedin}
If all of the gas in the IGM is cold before reionization and
subsequently jumps to $10^4$ K at
$a_{\rm rei}$, the filtering mass is \citep{gnedin98}
\begin{equation}
M_f = M_{\rm J} \, \left ( \frac{3}{10} \left[1 + 4 \left( \frac{a_{\rm
rei}}{a} \right)^{5/2} - 5 \left( \frac{a_{\rm rei}}{a}\right)^2
\right]\right)^{3/2}. \label{eq:fm}
\end{equation}
This mass scale can be much smaller than the Jeans mass for $10^4$ K
gas $M_{\rm J}$ and is typically time dependent. Since $M_f$ typically
corresponds to non-linear scales where linear theory is a poor
approximation, it is unclear how well equation (\ref{eq:fm})
represents the smallest mass at which the baryons clump. However,
\citet{gnedin98} smooth N-body simulations by including a pressure
force that becomes important at the filtering scale. They conclude
that this procedure reproduces well the small-scale gas power spectrum
seen in hydrodynamics simulations. Furthermore, \citet{gnedin00b}
finds that the filtering mass provides a good fit to the minimum
formation mass for a gas-rich halo.

\end{appendix}

\bibliographystyle{mn2e}

\end{document}